\documentclass[12pt]{iopart}
\usepackage{epsfig}
\usepackage{rotating}

\eqnobysec

\newcommand{\be}{\begin{equation}}
\newcommand{\ee}{\end{equation}}
\newcommand{\bea}{\begin{eqnarray}}
\newcommand{\eea}{\end{eqnarray}}
\newcommand{\beastar}{\begin{eqnarray*}}
\newcommand{\eeastar}{\end{eqnarray*}}
\newcommand{\nn}{\nonumber\\}
\newcommand{\lav}{\left\langle}
\newcommand{\rav}{\right\rangle}

\newcommand{\Tc}{T_{\rm c}}
\newcommand{\tw}{t_{\rm w}}
\newcommand{\yw}{y_{\rm w}}
\newcommand{\zw}{z_{\rm w}}
\newcommand{\uw}{u_{\rm w}}
\newcommand{\order}{{{\mathcal O}}}
\newcommand{\ie}{{\it i.e.}}
\newcommand{\eg}{{\it e.g.}}

\newcommand{\half}{\frac{1}{2}}
\newcommand{\rv}{\mathbf{r}}
\newcommand{\dz}{\Delta z}

\newcommand{\dX}{\Delta s}
\newcommand{\dt}{\Delta t}
\newcommand{\X}{s}
\newcommand{\dY}{\Delta r} 
\newcommand{\Om}{\Omega}
\newcommand{\om}{\omega}
\newcommand{\Y}{r}
\newcommand{\rn}{N^{-1/2}}
\newcommand{\eq}[1]{~(\ref{#1})}

\renewcommand{\sc}[1]{{\mathcal{F}}_{#1}}

\newcommand{\Inol}{\int dt' dt''\,}

\newcommand{\zv}{\mathbf{0}}
\newcommand{\qv}{\mathbf{q}}
\newcommand{\dq}{\int(dq)\,}
\newcommand{\dqp}{\int(dq')\,}
\newcommand{\dqd}{\int(dq'')\,}
\newcommand{\qpv}{\mathbf{q'}}
\newcommand{\qdv}{\mathbf{q''}}

\renewcommand{\sc}[1]{{\mathcal{F}}_{#1}}
\newcommand{\G}{{\mathcal{G}}}
\newcommand{\HH}{{\mathcal{H}}}
\newcommand{\Ltwo}{L^{(2)}}
\renewcommand{\eql}{_{\rm eq}}
\newcommand{\cd}{\frac{\lambda_d k_d}{4-d}}
\newcommand{\ccd}{\frac{\lambda_d k_d}{(4-d)(d-2)}}

\newcommand{\tbar}{\bar{t}}

\newcommand{\bth}{{\par\bf begin thesis}\\}
\newcommand{\eth}{{\par\bf end thesis}\\}

\begin{document}

\title[Dynamic heterogeneities in critical coarsening]{Dynamic
heterogeneities in critical coarsening: Exact results for correlation
and response fluctuations in finite-sized spherical models}

\author{Alessia Annibale\footnote[1]{Email
alessia.annibale@kcl.ac.uk}, Peter Sollich\footnote[2]{Email
peter.sollich@kcl.ac.uk}}

\address{King's College London, Department of Mathematics, London WC2R
2LS, UK}

\begin{abstract} 
We study dynamic heterogeneities in the out-of-equilibrium coarsening
dynamics of the spherical ferromagnet after a quench from
infinite temperature to its critical point.
A standard way of probing such heterogeneities is by
monitoring the fluctuations of correlation and susceptibility, coarse-grained 
over mesoscopic regions. We discuss how to define such fluctuating
coarse-grained correlations and susceptibilities in models where no quenched 
disorder is present. Our focus for the spherical model is on coarse-graining 
over the whole volume of $N$ spins, which requires accounting for
$\order(N^{-1/2})$ non-Gaussian fluctuations 
of the spin variables. The latter are treated as a 
perturbation about 
the leading order Gaussian statistics.
We obtain exact results for these quantities, which enable us to characterize 
the joint distribution of correlation and susceptibility fluctuations. 
We find that this distribution is qualitatively different, even for equilibrium above criticality, from the spin-glass scenario
where correlation and susceptibility fluctuations are linked in a manner 
akin to the fluctuation-dissipation relation between the average correlation
and susceptibility.
Our results show that coarsening at criticality is clearly heterogeneous 
above the upper critical dimension and suggest that, as in other glassy 
systems, there is a well-defined timescale on which fluctuations across 
thermal histories are largest. Surprisingly, however, neither this
timescale nor the amplitude of the heterogeneities increase with the age
of the system, as would be expected from the growing correlation length. Below
the upper critical dimension, the strength of 
correlation and susceptibility fluctuations varies on a timescale
proportional to the age of the system; the corresponding amplitude
also grows with age, but does not scale with the correlation volume as
might have been expected naively.
\end{abstract}

\section{Introduction}

Dynamic heterogeneities, where different local regions of a system evolve on 
different timescales, arise in many non-equilibrium situations.
Their existence
in supercooled liquids and other glassy systems has been probed experimentally 
using techniques including 
light scattering and confocal microscopy
\cite{WeeCroLevSchWei00,KegVan00,Ediger00,DesVan01,ThuEdi02,QiuEdi03} and this has been complemented by results from simulation and theory
\cite{YamOnu97,OnuYam98,YamOnu98,GloJanLooMacPoo98,FraDonParGlo99,Sear00b,CugIgu00,LacStaSchNovGlo02,Diezemann03}.

In systems with quenched disorder such as spin glasses, the disorder
itself provides an obvious source of heterogeneous dynamics: spins
compelled to assume particular local configurations by the disorder
decorrelate slowly, while less constrained ones can lose memory of
their configuration at some initial time very quickly. In the aging
dynamics of such systems after a quench to low temperature, it has
been argued that an invariance of the global dynamics under
reparametrization of time also dominates the fluctuations of the
local dynamics, with different regions of the system effectively
having different ages
\cite{CasChaCugKen02,ChaKenCasCug02,CasChaCugIguKen03,ChaCug07}.
In order to study such behaviour it is natural to consider two-time
correlation functions and susceptibilities. For a lattice system with
$N$ sites labelled by $i$ and spins (or more generally local order
parameters) $S_i$ these 
functions are, after spatial coarse-graining over the entire finite-sized
system\footnote{%
In $\hat C$ one should, in principle, subtract a term $m(t)m(\tw)$
with $m(t)=(1/N)\sum_i S_i(t)$, in order to get a connected
correlator. However, this subtracted term is $\mathcal{O}(1/N)$ in the scenarios
we consider and so contributes negligibly both to the $\mathcal{O}(1)$ average and
the leading $\mathcal{O}(N^{-1/2})$ fluctuation of $\hat{C}$.
},
\bea
\hat C(t,\tw) &=& \frac{1}{N}\sum_i\hat C_{ii}(t,\tw),
\quad \hat C_{ii}(t,\tw)=S_i(t)S_i(\tw)
\label{Ccg}
\\
\hat\chi(t,\tw) &=& \frac{1}{N}\sum_i\hat\chi_{ii}(t,\tw),
\quad \hat\chi_{ii}(t,\tw)=\frac{\partial S_i(t)}{\partial h_i(\tw)}\ .
\label{chicg}
\eea
Here $h_i$ is the field conjugate to $S_i$ and is assumed to have
been switched on at the waiting time $\tw$ (measured from the time of
preparation of the system, e.g.\ by quenching), with the response
measured at the later time $t$. We have used hats to emphasize that
the correlator and susceptibility defined above
fluctuate across thermal histories (including
variability in initial conditions); we return below to what this
implies for measuring $\hat\chi$. Averaging over the thermal
fluctuations gives the conventional correlation and susceptibility,
$C=\langle \hat C\rangle$ and $\chi=\langle\hat\chi\rangle$.  In aging
systems these are related by an out-of-equilibrium
fluctuation-dissipation (FD) relation~\cite{CugKur94}
\be
-\partial_{\tw}\chi(t,\tw)\equiv R(t,\tw)
=\frac{X(t,\tw)}{T}\partial_{\tw}C(t,\tw)
\label{FDT}
\ee
where $R(t,\tw)$ is the impulse response function, linked to the
susceptibility by
\be
\chi(t,\tw)=\int_{\tw}^t dt'\,R(t,t')
\label{integrated_chi}
\ee
and $X(t,\tw)$ is the fluctuation-dissipation ratio (FDR).  The FDR
can be read off from the negative slope of a parametric FD plot
showing $T\chi$ versus $C$, at fixed $t$ and with $\tw$ varying along
the curve.  In equilibrium, the FD theorem (see \eg~\cite{Reichl80})
implies that the FD plot is a straight line with slope $-1$,
corresponding to $X=1$. Out of equilibrium, the prediction from local time
reparametrization
invariance~\cite{CasChaCugKen02,ChaKenCasCug02,CasChaCugIguKen03,ChaCug07}
is that the contour lines of the joint probability distribution
of the {\em fluctuating} quantities $\hat C$ and $T\hat\chi$, at fixed $\tw$
and $t$, follow the local slope of the FD plot constructed from the
{\em average} quantities.

An obvious question to ask is whether the fluctuations of correlation
and response obey similar constraints in systems without quenched
disorder, where dynamic heterogeneities are ``self-generated''.  There
is some evidence for an affirmative answer from simulations of
kinetically constrained models of (structural)
glasses~\cite{ChaChaCugReiSel04}. Our interest here is in simpler
models displaying aging, where we can hope to make progress by
analytical calculation, namely coarsening
systems~\cite{MayBisBerCipGarSolTra04,MaySolBerGar05}. It has been argued
that in these full time-reparametrization invariance no longer holds,
with only time rescaling remaining as a symmetry in the long-time
aging dynamics~\cite{ChaCugYos05}. This means that there is no obvious
reason a priori for the presence of any constraint linking local
correlation and response fluctuations to the average FD relation, and
our main aim will be to investigate what effects this has on the
distributions of the fluctuating quantities.

Coarsening systems are magnetic systems -- or their analogues in
gas-liquid phase separation, demixing of binary liquids etc -- quenched from
the high-temperature phase to or below their critical temperature,
$\Tc$, (see \eg~\cite{GodLuc00b,MayBerGarSol03} and the review
\cite{CalGam05}).  
During the phase ordering (below $\Tc$) or the critical
relaxation (at $\Tc$), aging occurs  due to the
growth of a length scale (domain size or correlation length
respectively) \cite{Bray94}, and in an infinitely large system
equilibrium is never reached. At any time there are spins at the
interfaces that behave quite differently from the ones inside a
domain, so that the origin of dynamic heterogeneities and the
associated length scale have a clear physical interpretation. As
before one can then consider the resulting fluctuations around average
FD plots. For critical coarsening the analysis has to be modified
slightly: here the FD plots can in fact hide
the interesting aging behaviour of the FDR
$X$~\cite{GodLuc00b,MayBerGarSol03,SolFieMay02,CalGam04,GarSolPagRit05,AnnSol06,MaySol07}.
The latter typically is a smooth function of $\tw/t$~\cite{GodLuc00b},
so that also the fluctuation effects have to be analysed in terms
of the same scaling variable.

In this paper we study the fluctuations in the Langevin dynamics of
finite-size spherical ferromagnets \cite{BerKac52,Joyce72} after a
quench from equilibrium at infinite temperature to some low
temperature $T$. We will focus
mainly on quenches to the critical temperature, but comment also on
the behaviour in the equilibrium region above. Our calculation is
based on a leading order expansion in $1/N$ of the non-Gaussian
fluctuations of the spins, so that we are effectively considering
systems of finite but large size $N$. The nature of the expansion
prevents us from accessing the phase ordering below $\Tc$, where as
soon as domains are formed the non-Gaussian fluctuations become
dominant rather than a small perturbation about the leading order
statistics, which are Gaussian in the spherical model. Our results
therefore complement those of Ref.~\cite{ChaCugYos05}, where for the
case of zero temperature and $N\to\infty$ the fluctuations of correlations
were analysed, for coarse-graining volumes ranging from a single site
to much larger than the correlation length. The overall scaling
of correlation fluctuations in coarsening below $\Tc$ is 
well understood and has a clear interpretation in terms of 
the growing domain size~\cite{MayBisBerCipGarSolTra04}. Our study
fills an important gap in allowing us to look at coarsening {\em at}
criticality, where the connection between dynamical heterogeneities
and a growing correlation length is much less clear.

We will first describe, in Sec.~\ref{sec:definitions}, how we
characterize the fluctuations of coarse-grained correlation and
susceptibilities. A discussion is then given of possible alternative
definitions of these fluctuating quantities. Some of these can be
ruled out as less useful because they give different scalings with
system size for the variance of correlations and susceptibilities. 

Our analysis proper starts
in Sec.~\ref{sec:setup_hetero} with the derivation of the 
correlation and susceptibility variances and the covariance. We will 
give general, exact expressions for these quantities 
in terms of three-time kernels, $D$ and $D^{\chi}$, 
which will constitute the basis 
for all our further analysis. Our interest will be in the
out-of-equilibrium dynamics of the system after 
a quench from an initial state of equilibrium at high temperature to its 
critical temperature. Before considering non-equilibrium, though, 
we will analyse briefly the situation of a quench to above criticality, 
(in Sec.~\ref{sec:hetero_above_Tc}), where 
the equilibration process is fast and a genuine 
equilibrium dynamics takes place. We will first derive the general equilibrium 
expressions of the relevant quantities and later specilize to the case 
of high temperature. 
Even here, the results are new as far we know.
In Sec.~\ref{sec:hetero_dgt4} 
we will turn to the more interesting case of quenches to criticality. 
In the regime of small time 
differences the critical dynamics displayed by the system 
is stationary and one can look at the equilibrium situation.
For larger time differences the aymptotic dynamics in $d>4$ is essentially 
given by the equilibrium one
modulated by relatively weak aging corrections,
whereas  
for $d<4$ one needs to look directly at the non-equilibrium situation.
The latter is discussed in Sec.~\ref{sec:d_lt_4}. 
We summarize and look
forward to avenues for future research in Sec.~\ref{sec:discussion}.

\section{Definition of fluctuating correlation and susceptibility}
\label{sec:definitions}

In this section we describe first how we will characterize the
fluctuations of the correlator and susceptibility defined in
Eqs.\ (\ref{Ccg},\ref{chicg}). We then discuss, and largely rule out,
alternative ways of defining fluctuating correlation and response
functions that are coarse-grained across an entire finite system.

To the leading order in $1/N$, i.e.\ inverse system size, that we
keep, the joint distribution of the fluctuating
correlation and susceptibility is Gaussian and therefore fully
characterized by its second moments.  Specifically, denoting the
fluctuations $\delta C = \hat C - C$ and $\delta\chi = \hat \chi -
\chi$, we will study the behaviour of the variances
\bea
V_C(t,\tw) &=&N\lav[\delta C(t,\tw)]^2\rav
\label{variance_C}
\\
V_{\chi}(t,\tw)&=&NT^2\lav[\delta\chi(t,\tw)]^2\rav
\label{variance_R}
\eea
and of the covariance
\be
V_{C\chi}(t,\tw)=NT\lav\delta C(t,\tw)\delta\chi(t,\tw)\rav
\label{variance_cross}
\ee
Factors of $N$ have been included here to make all three quantities of
order unity. Also, as suggested by the equilibrium FD theorem, we have
scaled the susceptibility fluctuation $\delta\chi$ by a factor $T$ to
obtain a quantity with the same dimension as $\delta C$.

Writing out the variance of the fluctuations of the correlation explicitly as
\be
V_C(t,\tw) = \frac{1}{N}\sum_{i,j}\left[
\langle \hat C_{ii}(t,\tw) \hat C_{jj}(t,\tw)\rangle
- 
\langle \hat C_{ii}(t,\tw) \rangle \langle \hat C_{jj}(t,\tw)\rangle
\right]
\label{VC_chi4}
\ee
one sees that this is none other than the by now standard four-point
correlation function used to characterize heterogeneous
dynamics, often denoted $C_4$ or
$\chi_4$~\cite{FraDonParGlo99}. For coarsening below $\Tc$ in spatial
dimension $d$, the amplitude of this quantity scales with $\xi^d(\tw)
\sim \tw^{d/2}$~\cite{MayBisBerCipGarSolTra04}, where
$\xi(\tw)\sim\tw^{1/2}$ is the growing domain size. At criticality,
$\xi(\tw)$ -- now the lengthscale of regions across which equilibrium
correlations are established -- will still grow with the same exponent
but it is less obvious how it enters $V_C$ and the corresponding
susceptibility fluctuations. Our explicit results will shed light on
this question.

There are other fluctuating correlation and response functions we
could have considered. Firstly, instead of coarse-graining over the
entire finite-sized system as in\eq{Ccg} and\eq{chicg}, we could have
coarse-grained over regions of finite size (and then gathered
statistics also across all possible centre points of such
regions). However, for the spherical ferromagnet such a locally
coarse-grained susceptibility has negligible fluctuations compared to those
of the correlation, as demonstrated (in the context of the leading
order Gaussian spin statistics) in \cite{ChaCugYos05}. This is why we
focus on global coarse-graining, for which non-Gaussian
effects make also the response fluctuations non-trivial.

Secondly, our $\hat\chi$ from\eq{chicg} requires that we measure
separately, for a given noise history, all local susceptibilities
$\hat\chi_{ii}$. For Langevin dynamics as studied in this paper this
does not present a problem since the differentiation w.r.t.\ the local
field $h_i$ can be carried out directly, and an explicit equation of
motion for $\chi_{ii}$ be written down, in the spirit of a slave
estimator (see \eg~\cite{DeaDruHorMaj04}). However, already for Markov
dynamics simulated via a Monte Carlo scheme it becomes necessary in
principle to rerun the dynamics $N$ times, each time switching on one of the local
fields, unless specially crafted ``field-free'' methods are
used~\cite{Chatelain03,Ricci-Tersenghi03,Berthier07}. It is tempting to avoid this difficulty by using a standard trick for obtaining local
responses~\cite{Barrat98}: one could consider the observable
$A=\sum_i \epsilon_i S_i$, with the $\epsilon_i$ quenched zero mean
random variables. The response of $A$ to its conjugate field, scaled
by $1/N$ to give a result of order unity, is then
\be
\hat\chi_\epsilon = \frac{1}{N} \sum_{ij} \epsilon_i \epsilon_j \hat\chi_{ij}.
\label{hat_chi_epsilon}
\ee
It can be measured with a single rerun of the history (although of
course even this is likely to be impossible in a real rather than a
numerical experiment). The above procedure, employed
in~\cite{CasChaCugKen02,CasChaCugIguKen03}, should reduce to the
definition\eq{chicg} if the random field amplitudes $\epsilon_i$ are
drawn without spatial correlation so that, on averaging over their
distribution as indicated by the overbar,
$\overline{\epsilon_i\epsilon_j}=\delta_{ij}$. One has to bear in
mind, however, that the randomness in the $\epsilon_i$ may induce
additional fluctuations in $\hat\chi_\epsilon$ which are not present
in $\hat\chi$. This does not appear to have been the case for the
systems studied in~\cite{CasChaCugKen02,CasChaCugIguKen03}, presumably
due to the presence there of quenched disorder.

In our case, on the other hand, the variance of $\hat\chi_\epsilon$
would be genuinely larger than that of $\hat\chi$. This can be seen by
considering
\be
\fl N\langle (\delta \chi_\epsilon)^2\rangle =
\frac{1}{N}\sum_{ijkl}\epsilon_i\epsilon_j\epsilon_k\epsilon_l
\lav \delta \chi_{ij}\delta \chi_{kl}\rav =
\sum_{j'k'l'}
\left(\frac{1}{N}\sum_i\epsilon_i\epsilon_{i+j'}
\epsilon_{i+k'}\epsilon_{i+l'}\right)
\lav \delta \chi_{0j'}\delta \chi_{k'l'}\rav 
\ee
where we have used that after thermal averaging the response
statistics in a system without quenched disorder are translationally
invariant. As the statistics of the $\epsilon$ are defined to be likewise
translationally invariant, the normalized sum over $i$ can be replaced
to leading order by a disorder average, giving
\be
N\langle (\delta \chi_\epsilon)^2\rangle =
\sum_{j'k'l'}
\overline{\epsilon_0\epsilon_{j'}\epsilon_{k'}\epsilon_{l'}}
\lav \delta \chi_{0j'}\delta \chi_{k'l'}\rav 
=
\frac{1}{N}\sum_{ijkl}\overline{\epsilon_i\epsilon_j\epsilon_k\epsilon_l}
\lav \delta \chi_{ij}\delta \chi_{kl}\rav
\label{chi_eps_variance}
\ee
The fourth-order disorder average gives 
\be
\overline{\epsilon_i\epsilon_j\epsilon_k\epsilon_l}=
\overline{\epsilon_i\epsilon_j}\ \overline{\epsilon_k\epsilon_l}+
\overline{\epsilon_i\epsilon_k}\ \overline{\epsilon_j\epsilon_l}+
\overline{\epsilon_i\epsilon_l}\ \overline{\epsilon_j\epsilon_k}
\ee
This is exactly true
if the $\epsilon$ are taken as Gaussian variables; for \eg\ binary
variables $\epsilon_i=\pm 1$ one gets an extra term
$-2\delta_{ij}\delta_{ik}\delta_{il}$ but this makes a subleading
contribution in $1/N$. Overall one has
\be
N\langle (\delta \chi_\epsilon)^2\rangle =
\frac{1}{N}\sum_{ij}\left[\langle \delta \chi_{ii}\delta \chi_{jj}\rangle 
+ 2 \langle (\delta \chi_{ij})^2\rangle\right]
\label{chi_eps_variance_final}
\ee
Comparing with\eq{chicg} one sees that this is indeed larger than
$N\langle(\delta\chi)^2\rangle$, by the second term
in the square brackets of\eq{chi_eps_variance_final}. As an aside, we
note that\eq{chi_eps_variance} can be written as $\langle (\delta
\chi_\epsilon)^2\rangle =
\overline{\langle (\delta \chi_\epsilon)^2\rangle}$, \ie\ the variance
of $\delta\chi_\epsilon$ is self-averaging with respect to the
sampling of the field amplitudes $\epsilon$. The same argument can be
applied to all other moments of $\hat\chi_\epsilon$, so that the
entire distribution $P(\hat\chi_\epsilon)$ is self-averaging. The
increased variance of $\hat\chi_\epsilon$ is therefore present for any
given sample of the $\epsilon$. It does not arise, as one might
alternatively have suspected, by $\hat\chi_\epsilon$ for each
sample $\epsilon$ having a distribution similar to that of $\hat\chi$ but with
a shifted mean that fluctuates with $\epsilon$.

The difference between $\hat\chi_\epsilon$ and $\hat\chi$ can be
avoided by averaging over a sufficiently large number of different
configurations of the $\epsilon_i$. (This of course means that
an appropriate number of reruns of each thermal history are required,
defeating to a certain extent the object of working with the random field
amplitudes $\epsilon_i$.) One thus
effectively ``preaverages'' over $\epsilon$; allowing for a general
covariance $\overline{\epsilon_i\epsilon_j}\equiv\epsilon_{ij}$ this
gives
\be
\hat\chi = \frac{1}{N}\sum_{ij}\epsilon_{ij}\hat\chi_{ij}
\label{staggered_chi}
\ee
It is this form that we will use in the calculations below, with
$\epsilon_{ij}$ short-ranged (so that $\sum_j
\epsilon_{ij}=\mathcal{O}(1)$). The extreme long-range case $\epsilon_{ij}=1$
corresponds to spatially uniform, non-disordered, fields
$\epsilon_i=1$ and so would be easiest to measure, with only a single
rerun of the thermal history. The observable $A$ then simplifies to
$A=\sum_i S_i=Nm$ so that $\hat\chi$ becomes the magnetization
susceptibility $\hat\chi_m$. Its (scaled) variance can be written as
\be
N \langle(\delta\chi_m)^2\rangle =
\frac{1}{N}\sum_{ijkl}\langle\delta\chi_{ij}\delta\chi_{kl}\rangle
\label{chi_m_var}
\ee
Significant contributions to the sum are expected to arise only when
all sites $i$, $j$, $k$ and $l$ are close to each other spatially
($i$, $j$ must not be too far apart to give a sizable response at all,
similarly for $k,l$, and then these two pairs of sites need to be
close to each other to have correlated response fluctuations), giving an
$\mathcal{O}(1)$ result as for the other susceptibilities considered so far.

The reason why we will not consider $\hat\chi_m$ further is that the
corresponding correlation function $\hat C_m = (1/N)\sum_{ij} \hat
C_{ij}$ has a variance that is much larger, by a factor of order $N$.
To see this, write $\hat C_m(t,\tw) = N m(t)m(\tw)$ and consider the simplest
case of the equal-time correlation $\hat
C_m(t,t)=[\sqrt{N}m(t)]^2$. The magnetization has fluctuations of
order $1/\sqrt{N}$ around zero, so that $\sqrt{N}m(t)$ has zero mean
and fluctuations of order unity (or, more precisely, of order
$\xi^{d/2}(t)$ in $d$ dimensions). The correlation function $\hat
C_m(t,t)=[\sqrt{N}m(t)]^2$ is therefore also of order unity but,
crucially, has fluctuations of the {\em same} order. It follows that
$N\langle (\delta C_m)^2\rangle$ is of order $N$ as claimed. The same
argument applies to $\hat C_\epsilon$ defined in analogy
with\eq{hat_chi_epsilon}: one writes $\hat
C_\epsilon(t,\tw)=(1/N)\sum_{ij} \epsilon_i \epsilon_j \hat
C_{ij}(t,\tw) = N[A(t)/N][A(\tw)/N]$ with the staggered magnetization
$A(t)/N = (1/N)\sum_i \epsilon_i S_i(t)$ which scales in the same way
as $m(t)$.

One can phrase the argument for these large correlation fluctuations
differently, to see more clearly where the difference to the
susceptibility fluctuations arises. Taking the magnetization
correlator, one has by analogy with\eq{chi_m_var}
\be
N \langle(\delta C_m)^2\rangle =
\frac{1}{N}\sum_{ijkl}\langle\delta C_{ij}\delta C_{kl}\rangle
\label{C_m_var}
\ee
When the sites $i$, $j$ are far apart, $C_{ij}$ is small and so
$\delta C_{ij} = \hat{C}_{ij}-C_{ij}\approx
\hat{C}_{ij}=S_i(t)S_j(\tw)$. But then $\delta C_{ij}\delta C_{kl}
\approx S_i(t)S_j(\tw)S_k(t)S_l(\tw)$ can still be substantial as long as
$i,k$ are close and similarly $j,l$ (or $i,l$ and $j,k$). There are
$\order(N^2)$ such terms in the sum\eq{C_m_var}, giving a scaled
variance of $\hat C_m$ of $\order(N)$ as claimed. The same argument
can be applied to the variance of $C_\epsilon$ for uncorrelated
$\epsilon_i$, which is given by an expression analogous
to\eq{chi_eps_variance_final}.

Only by preaveraging over the field amplitudes $\epsilon_i$ does one
obtain a correlation function with fluctuations of the same order as
the corresponding susceptibility. By analogy with\eq{staggered_chi},
this correlator can be written as
\be
\hat C = \frac{1}{N}\sum_{ij}\epsilon_{ij}\hat C_{ij}
\label{staggered_C}
\ee

In summary, the only sensible definitions of the fluctuating
correlation and susceptibility that involve coarse-graining across the
entire system appear to be\eq{staggered_C} and\eq{staggered_chi};
other definitions involving quenched field amplitudes $\epsilon_i$
without preaveraging lead to correlation variances that are larger
than those of the susceptibility by a factor of $\order(N)$. The
arguments we have given apply quite generically for systems without
quenched disorder. \footnote{They apply also to coarse-graining over 
finite volumes, as long as we
are considering  moderate timescales where the typical correlation
volume remains much smaller than the coarse-graining volume: again the
alternative definitions that we have considered would give a correlation
variance much larger than the susceptibility variance, by a factor of
the order the ratio of coarse-graining volume to correlation volume.
}
In the spherical model the situation is, in fact,
somewhat more complicated because of the effective long-range
interaction between spins arising from the spherical constraint. The
resulting weak but long-range correlations lead to extra contributions
to the fluctuations of $\hat C$ but without changing the scaling with
$N$; for the susceptibility, these long-range terms provide the {\em
only} source of fluctuations but again the scaling with $N$ 
is unaffected.

We will retain the preaveraged field correlations $\epsilon_{ij}$ as
essentially arbitrary short-ranged quantities during the initial part
of our analysis, but then simplify in the concrete evaluation to the
case of coarse-grained local quantities,
$\epsilon_{ij}=\delta_{ij}$, effectively returning to the
definitions (\ref{Ccg},\ref{chicg}) given in the
introduction. Investigation of the more general case 
could be an interesting subject of future work; indeed, only for zero
temperature, where spins within domains are fully correlated with each
other, would one expect to obtain correlation fluctuations
equivalent to those for the local case.

\section{Setup of calculation}
\label{sec:setup_hetero}

We analyse the mesoscopic fluctuations in the dynamics of the spherical 
ferromagnet
\be
H = \half \sum_{(ij)} (S_i-S_j)^2
\label{H_spherical}
\ee
where the sum runs over all nearest neighbour (n.n.) pairs on a
$d$-dimensional unitary (hyper-)cubic lattice. 
The spins $S_i$ are real variables at each of the $N$ lattice 
sites $\rv_i$, subject to the spherical constraint $\sum_i S_i^2=N$.

The Langevin equation for this system subject to thermal noise $\xi_i$ 
can be written as~\cite{AnnSol06}
\be
\partial_t S_i = - \frac{\partial H}{\partial S_j} +\xi_i - (z_0(t)+ N^{-1/2}
 z_1(t)) S_i
\label{langevein0}
\ee
where $z_0(t)$ is the Lagrange multiplier implementing the spherical 
constraint
and $N^{-1/2}z_1$ is its leading fluctuation of
$\order{(N^{-1/2})}$. The latter
is conventionally neglected in the Gaussian theory, and this is
justified for observables that probe correlations on scales 
small compared to the size of the system. For globally coarse-grained
quantities like our $\hat C$ and $\hat\chi$,
on the other hand, one requires the correlations of
all the spins of the system. The fluctuations of $\order{(N^{-1/2})}$
are then no longer negligible and the Gaussian theory becomes invalid
\cite{AnnSol06}.
One can also write\eq{langevein0} in terms of the discrete (lattice) Laplacian $\Omega$,
which takes the values $\Omega_{ii}=2d$ on the diagonal and $\Omega_{ij}=-1$ 
for n.n.\ sites $i,j$:
\be
\partial_t S_i = - \sum_j \Omega_{ij} S_j +\xi_i - (z_0(t)+ N^{-1/2}
 z_1(t)) S_i
\label{langevein1}
\ee
One expects that the fluctuations in the Lagrange multiplier of 
$\order{(N^{-1/2})}$ induce non-Gaussian fluctuations in the spin variables 
of the same order. To account for this we decompose
the spin variables as
$S_i = \X_i + N^{-1/2} \Y_i$, where $\X_i$ gives the limiting
result for $N\to\infty$, which has purely Gaussian statistics, and
$N^{-1/2}\Y_i$ is the leading-order non-Gaussian fluctuation correction.   
Inserting this decomposition into\eq{langevein1} and collecting terms of 
$\order{(1)}$ and $\order{(N^{-1/2})}$ gives
\bea
\partial_t s_i = - \sum_j \Omega_{ij} s_j +\xi_i - z_0(t)s_i
\label{lang_gaussian}
\\
\partial_t r_i = - \sum_j \Omega_{ij} r_j - z_0(t)r_i -z_1(t)s_i
\label{lang_correction}
\eea
In terms of the Fourier components $S_\qv = \sum_i s_i
\exp(-i\qv\cdot\rv_i)$ 
of the spins the Gaussian dynamics\eq{lang_gaussian} reads 
\be
\partial_t S_\qv = -(\omega_\qv+z_0(t))S_\qv + \xi_\qv
\label{dotSq}
\ee
where $\omega_\qv = 2\sum_{a=1}^d (1-\cos q_a)$.
Its solution with initial condition at time $\tw$ is 
\be
S_\qv(t) = R_\qv(t,\tw) S_\qv(\tw) + \int_{\tw}^t dt'\,R_\qv(t,t')\xi_\qv(t')
\label{Sqt}
\ee
given in terms of the two-time Fourier mode response function 
\be
R_\qv(t,\tw) = \exp\left(-\omega(t-\tw)-\int_{\tw}^t dt'\,
z(t)\right) \equiv \sqrt{\frac{g(\tw)}{g(t)}}e^{-\omega(t-\tw)}
\label{Rq}
\ee
where the subscript $\qv$ in $\omega_{\qv}$ has been omitted and
\be
g(t) = \exp\left(2\int_{0}^t dt'\, z_0(t')\right).
\label{g_def}
\ee
The two-time correlator in the Gaussain theory reads 
$C_\qv(t,\tw) = (1/N) \lav
S_\qv(t)S_\qv^*(\tw)\rav$ and follows from\eq{Sqt} by
propagating the equal-time 
correlator $C_\qv(\tw,\tw) = (1/N) \lav
S_\qv(\tw)S_\qv^*(\tw)\rav$ from initial time 
$\tw$ to final time $t$
\be
C_\qv(t,\tw) = R_\qv(t,\tw) C_\qv(\tw,\tw)
\label{C_twotime}
\ee
Once the function $g(t)$ is known, these results capture all of the
leading order Gaussian dynamics of the spins. Notice that the impulse
reponse\eq{Rq} is deterministic: there are no response fluctuations
within the Gaussian theory.

To determine the non-Gaussian corrections\eq{lang_correction}, one needs to
have an expression for the Lagrange multiplier fluctuations. This can
be worked out as~\cite{AnnSol06}
\be
z_1=\half\int dt'\,L(t,t')\Delta(t')
\label{z1}
\ee
where $\Delta(t)$ is an $\order{(1)}$ quantity
describing the fluctuations of the squared length of the Gaussian spin
variables $s_i$ 
\be
\Delta(t)=\frac{1}{\sqrt{N}}\sum_l(s_l^2(t)-1)
\label{delta}
\ee
and $L$ is the inverse operator of the kernel $K$
\be
K(t,t')=\frac{1}{N}\lav s_i(t)R_{im}(t,t')s_m(t')\rav=
\frac{1}{N}R_{im}(t,t')C_{im}(t,t')
\label{K_definition}
\ee
defined by
\be
\int dt' K(t,t')L(t',\tw)=\delta(t-\tw)
\label{L_definition}
\ee
Both $K$ and $L$ are causal, \ie\ they vanish for $t<\tw$.
Above and in what follows the summation convention for 
repeated indices is used. 
In\eq{K_definition},\, $R_{im}$ is the inverse Fourier transform of\eq{Rq}, 
$R_{im}=(1/N)\sum_\qv e^{i\qv\cdot(\mathbf{r}_i-\mathbf{r}_m)}R_{\qv}$, with 
the sum running over the $N$ appropriate wavevectors $\qv$; for even
$L$ their components
are integers in the range $-L/2\ldots -1,0,1,\ldots L/2-1$
multiplied by an overall factor $2\pi/L$. 
When considering continuous functions of $\qv$ this sum
can be replaced by the integral $\dq$, 
where we abbreviate $(dq) \equiv
d\qv/(2\pi)^d$, and the integral runs over the first Brillouin zone
of the hypercubic lattice, i.e.\ $\qv\in[-\pi,\pi]^d$; this
simplification will apply throughout 
our analysis. In Fourier space the kernel\eq{K_definition} then reads
\be
K(t,t')=
\dq R_{\qv}(t,t')C_{\qv}(t,t')
\label{Kq_definition}
\ee
The non-Gaussian corrections to the spins are determined 
by solving the dynamical equation\eq{lang_correction}, and can be expressed in terms of the Gaussian spins as
\be
r_i(t)=-\half\int dt'dt''\, R_{ik}(t,t')s_k(t')L(t',t'')\Delta(t'')
\label{ri}
\ee

As explained in the introduction, the object of our study are the globally 
coarse grained correlation and susceptibility functions, 
\bea
\hat C(t,\tw)=\frac{1}{N}\sum_{ij}\epsilon_{ij}S_i(t)S_j(\tw)
\label{C}
\\
\hat\chi(t,\tw)=\frac{1}{N}\sum_{ij}\epsilon_{ij} 
\frac{\partial S_i(t)}{\partial h_j(\tw)}
\label{chi}
\eea
For the correlation function we insert the spin decomposition
$S_i=s_i+r_i/\sqrt{N}$ and expand to the order $1/\sqrt{N}$ of the
fluctuations we are interested in:
\be
\hat C(t,\tw)=\frac{1}{N}\sum_{ij}\epsilon_{ij}\left[s_i(t)
s_j(\tw)+\frac{1}{\sqrt{N}}\left(r_i(t)s_j(\tw)+
s_i(t)r_j(\tw)\right)\right]
\label{hatC}
\ee

To obtain the corresponding susceptibility we need
to expand the spin variables in both the magnetic field {\it and} $N^{-1/2}$.
More specifically, consider perturbing the system by an external field 
${h_i}=h\epsilon_i$ that couples linearly to the spins ${S_i}$. We
keep the $\epsilon_i$ fixed initially and perform the preaveraging afterwards.
The equation of motion in presence of the perturbation reads 
\be
\partial_t S_i = -\Om_{ij}S_j - \left(z_0(t)+\frac{z_1(t)
+h\dz_1(t)}{N^{1/2}}\right)S_i+h_i+\xi_i
\label{LangPert}
\ee
where now a change in the Lagrange multiplier induced by the field
perturbation, $N^{-1/2}h\dz_1$, is present in addition to
the fluctuating component $z_1$ of $\order{(\rn)}$ of the unperturbed
dynamics.
(One can show that there is no $\order(h)$ perturbation in the
Lagrange multiplier; such a term appears only if the system has a
finite magnetization and is perturbed by a uniform field~\cite{AnnSol06}.)
Inserting the corresponding expansion for the spin variables
\be
S_i=\X_i+h\dX_i+\frac{\Y_i+h\dY_i}{N^{1/2}}
\label{spin_expansion}
\ee
and collecting the $\order{(N^0)}$ terms gives 
to $\order{(h^0)}$ the unperturbed equation of motion for $\X_i$ and to 
$\order{(h^1)}$ a deterministic equation for 
the perturbed components 
\be
\partial_t\dX_i = -\Om_{ij}\dX_j - z_0(t)\dX_i +\epsilon_i
\ee
Integrated in time with the condition $\dX_i(t)=0$ for $t<\tw$ this gives 
\be
\dX_i(t)=\chi_{ij}(t,\tw)\epsilon_j
\label{dot_dsi}
\ee
where $\chi_{ij}$ is the non-fluctuating Gaussian susceptibility
$\chi_{ij}(t,\tw)=\int_{\tw}^t dt'\, R_{ij}$.

Gathering the 
$\order{(N^{-1/2})}$ terms in\eq{LangPert}, on the other hand, gives to 
$\order{(h^0)}$ equation\eq{lang_correction}, as expected, and to 
$\order{(h^1)}$ a new equation
\be
\partial_t\dY_i = -\Om_{ij}\dY_j - z_0(t)\dY_i -\dz_1(t)\X_i-z_1\dX_i
\label{dot_dri}
\ee
for the $\dY_i$, with solution
\be
\dY_i(t)=-\int_{\tw}^tdt'\,R_{ij}(t,t')[\dz_1(t')s_j(t')+z_1(t')\dX_j(t')]
\label{dri}.
\ee
The fluctuations in the response thus arise from the fluctuations of the 
Lagrange parameter, as anticipated. 
The $\Delta z_1$ term can be worked out by imposing 
that, due to the spherical constraint, $N^{-1}\sum_iS_i^2(t)=1$ at all times. 
Using\eq{spin_expansion}, this implies that the quantity
\bea
\fl\frac{1}{N}\sum_i (S_i^2-1)&=&\frac{1}{N}\sum_i\left(s_i^2-1
+2s_i \frac{r_i+h\dY_i}{\sqrt{N}}+2h\dX_i \,\X_i 
+ 2h\dX_i\frac{r_i}{\sqrt{N}}\right)
\\
\fl&=&\frac{\Delta}{\sqrt{N}}+2\frac{1}{N^{3/2}}\sum_i s_i\Y_i
+\frac{2h}{N}\sum_is_i\dX_i+\frac{2h}{N^{3/2}}\sum_is_i\dY_i
\nn
&&+\frac{2h}{N^{3/2}}\sum_ir_i\dX_i
\label{length0}
\eea
must vanish to the leading order in $h$, $N^{-1/2}$ and 
$h N^{-1/2}$; we have temporarily re-instated the summation
signs for clarity. To make progress, let us note that the first two 
terms on the r.h.s.\ of\eq{length0} cancel to $\order{(N^{-1/2})}$;
this is in fact how the corrections $r_i$ are determined~\cite{AnnSol06}.
In the third and fifth term we can insert the deterministic quantities
$\dX_i$ from\eq{dot_dsi}.
Since the $r_i$ are ``driven'' by the $s_i$ according to\eq{ri}, they
will only have spatial correlations of finite range. Thus 
$(1/N)\sum_i r_i\dX_i$  in the fifth term 
is $\order{(N^{-1/2})}$, making this
contribution $\order(hN^{-1})$ overall and subleading compared to the 
third term, which is $\order{(hN^{-1/2})}$. So we need to impose
\be
\frac{1}{\sqrt{N}}\sum_i s_i\left(\dX_i+\frac{\dY_i}{\sqrt{N}}\right)=0
\ee
to $\order{(1)}$, which yields using\eq{dri}
\bea
\fl\frac{1}{N}\int_{\tw}^t dt'\,R_{im}(t,t')s_i(t)s_m(t')\dz_1(t')&=&
\frac{1}{\sqrt{N}}\,s_i(t)\dX_i(t)
\nn
&&-\frac{1}{N}\int_{\tw}^tdt'\,R_{im}(t,t')s_i(t)\dX_m(t')z_1(t')
\label{eq_4_dz1}
\eea
In the second term on the RHS,
$R_{im}(t,t')\dX_m(t')=R_{im}(t,t')\chi_{mj}(t',\tw)\epsilon_j$ is a
deterministic $\order{(1)}$
quantity which is then summed over sites $i$ multiplied by the
short-range correlated $s_i(t)$. Together with the $1/N$ prefactor this
gives a negligible $\order{(N^{-1/2})}$ contribution.
On the LHS, $(1/N)R_{im}(t,t')s_i(t)s_m(t')$ has fluctuations of
$\order(N^{-1/2})$ which can likewise be neglected compared to its 
$\order{(1)}$ average; the latter equals $K(t,t')$ from\eq{Kq_definition}.
Inverting the resulting convolution $\int_{\tw}^t dt'\,K(t,t')\dz_1(t')$
using\eq{L_definition} one finds as the solution of\eq{eq_4_dz1} 
\be
\dz_1(t')=\frac{1}{\sqrt{N}}\int dt''\, L(t',t'')s_i(t'')\dX_i(t'')
\ee
With this we can now write down the susceptibility for
the given set of $\epsilon_i$, as defined in\eq{hat_chi_epsilon}. Noticing that
$\hat\chi_{ij}\epsilon_j$ is the response of spin $i$, given by the
$\order(h)$ terms from\eq{spin_expansion}, one gets
\bea
\fl\hat\chi_\epsilon(t,\tw)&=&\frac{1}{N}\epsilon_i\left(\dX_i+
\frac{1}{\sqrt{N}}\dY_i\right)\nn
\fl&=&\frac{1}{N} \epsilon_i\epsilon_j\chi_{ij}(t,\tw)
- N^{-3/2}\epsilon_i\int_{\tw}^t dt'\,
R_{im}(t,t')\times
\nn
\fl&&\times
\Biggl[\int dt''\,s_m(t')L(t',t'')\frac{1}{\sqrt{N}}s_n(t'')
\dX_n(t'')+\dX_m(t')z_1(t')\Biggr]
\eea
The first term is the non-fluctuating Gaussian contribution. The
fluctuating remainder becomes, once we insert\eq{dot_dsi} and
preaverage over the $\epsilon_i$, 
\bea
\fl\delta\chi(t,\tw)&=&
-N^{-2}\epsilon_{ij}\int_{\tw}^t dt' dt''\,R_{im}(t,t')
s_m(t')L(t',t'')s_n(t'')\chi_{nj}(t'',\tw)\nn
\fl &&{}- N^{-3/2}\epsilon_{ij}\int_{\tw}^t dt'\,R_{im}(t,t')\chi_{mj}(t',\tw)z_1(t')
\label{delta_chi}
\eea
In this expression the first term 
is $\order{(1/N)}$: the 
sum over $i,j$ gives an $\order{(1)}$ translation
invariant function 
$\epsilon_{ij} R_{im}(t,t')\chi_{nj}(t',\tw)$, and $s_m(t')s_n(t'')$
can be replaced by its average $C_{mn}(t',t'')$ to leading order; with
the $1/N^2$ prefactor and the summation over $m,n$ one gets
$\order(1/N)$ overall. (The neglected fluctuations of
$s_m(t')s_n(t'')$ will give an even smaller correction, of
$\order{(N^{-3/2})}$.) In the second term one argues similarly that
the sum of $(1/N)\epsilon_{ij}R_{im}\chi_{mj}$ over $i,j,m$ is $\order(1)$.
Since $z_1$ is scaled to be 
$\order{(1)}$, it is then this term that provides the leading
susceptibility fluctuation of $\order{(N^{-1/2})}$. 
Inserting\eq{z1} 
and\eq{delta} we can finally write
\be
\fl \delta\chi(t,\tw)=-\frac{1}{2} N^{-2}\epsilon_{ij}\int dt' dt''\,R_{im}(t,t')
L(t',t'')\sum_n(s_n^2(t'')-1)\chi_{mj}(t'',\tw)
\ee
We have dropped the integration limits since these are enforced
automatically by causality of $R_{im}$, $L$ and $\chi_{mj}$.

In order to study the fluctuations of globally coarse-grained quantities 
around their mean values, we will consider their variances and
covariance, defined in\eq{variance_C}, \eq{variance_R} and\eq{variance_cross}.
For the correlation variance one has, by inserting\eq{hatC}
into\eq{variance_C} and multiplying out,
\bea
\fl V_C(t,\tw)&=&
\frac{1}{N}
\epsilon_{ij}\,
\epsilon_{kl}\left\{\lav s_i(t)s_j(\tw)s_k(t)s_l(\tw)\rav'\right.
\nn
\fl&&{}+\frac{1}{\sqrt{N}}
\left[\lav r_i(t)s_j(\tw)s_k(t)s_l(\tw)\rav'
+\lav s_i(t)r_j(\tw)s_k(t)s_l(\tw)\rav'\right.
\nn
\fl&&{}+\left.\lav s_i(t)s_j(\tw)r_k(t)s_l(\tw)\rav'
+\lav s_i(t)s_j(\tw)s_k(t)r_l(\tw)\rav'\right]
\nn
\fl {}&&{}+\frac{1}{N}\left[
\lav r_i(t)s_j(\tw)r_k(t)s_l(\tw)\rav' 
+\lav r_i(t)s_j(\tw)
s_k(t)r_l(\tw)\rav'\right.
\nn
\fl &&
{}+\left.\left.\lav s_i(t)r_j(\tw)r_k(t)s_l(\tw)\rav'
+\lav s_i(t)r_j(\tw)s_k(t)r_l(\tw)\rav'\right]\right\}
\label{VC}
\eea
where the prime on the averages indicates that the corresponding disconnected
contributions arising from $\langle \hat C_{ij}\rangle\langle
\hat C_{kl}\rangle$ are to be subtracted. The susceptibility
variance\eq{variance_R} reads
\bea
\fl V_{\chi}(t,\tw)&=&
\frac{T^2}{4N^3}\epsilon_{ij}\,
\epsilon_{kl}\lav\int_{\tw}^t dt' dt'' d\tw' d\tw''\,R_{im}(t,t')L(t',t'')
R_{kp}(t,\tw')L(\tw',\tw'')\right.
\nn
\fl &&\times
\left.
\sum_n
(s_n^2(t'')-1)\sum_r(s_{r}^2(\tw'')-1)\chi_{mj}(t',\tw)\chi_{pl}(\tw',\tw)\rav
\label{VR}
\eea
%
while for the cross correlation\eq{variance_cross}, one has
\bea
\fl V_{C\chi}(t,\tw)&=&
-\frac{T}{2N^2}\epsilon_{ij}\,
\epsilon_{kl}\int_{\tw}^t dt' dt''
L(t',t'')R_{kp}(t,t')\chi_{pl}(t',\tw)
\nn
\fl &&\times\lav \left(s_i(t)+\frac{1}{\sqrt{N}}r_i(t)\right)\left(s_j(\tw)+\frac{1}{\sqrt{N}}r_j(\tw)\right)
\sum_n(s_n^2(t'')-1)\rav'
\label{VCR}
\eea
Since all the quantities appearing in the averages can be expressed,
via\eq{ri}, in terms of Gaussian variables $s_i$, we can use Wick's
theorem to perform the averaging. This gives a sum over all possible pairings
of the Gaussian variables, each contributing a product
of correlation functions. In the primed averages in $V_C$, pairings 
that do not couple the index groups $[ij]$ and $[kl]$ need to be
discarded, and similarly in $V_{C\chi}$. Fortunately, many other pairings
can also be dropped because they give subleading terms in $1/N$.
We omit the details as the reasoning is analogous to that
in~\cite{AnnSol06}, and state the results only for the coarse-grained
local correlation and susceptibility ($\epsilon_{ij}=\delta_{ij}$).

In order to make the expressions more manageable let us define 
\be
CC(t,\tw)=\dq C_{\qv}^2(t,\tw)
\label{CC2times}
\ee
and the following three-time function (we use the same symbol as the
number of arguments will make it clear which function is meant; note
that $CC(t,t,t')=CC(t,t')$)
\bea
CC(t,\tw,t')=\dq C_{\qv}(t,t')C_{\qv}(\tw,t')
\label{CC3times}
\eea
We also introduce
\be
\tilde{D}(t_1,t_2,t')=\int dt'' L(t'',t')\dq R_{\qv}(t_1,t'')C_{\qv}
(t_2,t'')
\label{Dtilde}
\ee
as well as
\be
D(t,\tw,t')=\half\left[\tilde{D}(t,\tw,t')+\tilde{D}(\tw,t,t')\right]
\label{D}
\ee
which is the symmetrized version of\eq{Dtilde}. Note that
$D(t,\tw,t')$ is causal in the sense that it vanishes for $t'>t$.
In terms of these functions the correlation
variance
takes the compact form 
\bea
\fl V_C(t,\tw)&=&\dq C_{\qv}(t,t)C_{\qv}(\tw,\tw)+CC(t,\tw)-4\int dt'\,D(t,\tw,t')CC(t,\tw,t')\nn
\fl &&{}+2\int dt'd\tw' D(t,\tw,t')D(t,\tw,\tw')CC(t',\tw')
\label{VC_D}
\eea
Similarly one can define
\be
D^{\chi}(t,\tw,t')=T\int dt''\,L(t'',t')\dq R_{\qv}(t,t'') \chi_{\qv}(t'',\tw)
\label{Dchi}
\ee
and express the susceptibility
variance
as
\be
V_{\chi}(t,\tw)=\half\int dt'd\tw'\,D^{\chi}(t,\tw,t')D^{\chi}(t,\tw,\tw')CC(t',\tw')
\label{VR_D}
\ee
In the susceptibility the times are already ordered and we do not need to consider
a symmetrized version of $D^{\chi}$; $D^\chi$ is causal in the same
sense as $D$.
The
covariance, finally,
can be expressed in terms of the same functions as
\bea
\fl V_{C\chi}(t,\tw)&=&-\int dt' D^{\chi}(t,\tw,t')CC(t,\tw,t')
\nn
& &{}+
\int dt'd\tw' D(t,\tw,t')D^{\chi}(t,\tw,\tw')CC(t',\tw')
\label{Vcross_D}
\eea
All the properties of the (co)variances can now be obtained from the
behaviour of the functions $CC$, $D$ and $D^{\chi}$. To understand the
general structure of $D$ and $D^\chi$  we first recall~\cite{AnnSol06}
that $L$, the inverse kernel of $K$, has the from
\be
L(t,t')=\delta'(t-t')+2T\delta(t-t')-\Ltwo(t,t')
\label{L_pieces}
\ee
where $\Ltwo(t,t')$ vanishes for $t'>t$, has a jump discontinuity at
$t'=t$ and is expected to be smooth and positive for $t'<t$. The
singular terms are consequences of the fact that
$K(t,t')$ vanishes for $t'>t$ and has equal-time value and slope
\be
\lim_{t\rightarrow t'^+}K(t,t')=1, \qquad
\partial_{t'}K(t,t')|_{t=t'^+}=2T
\label{properties_K} 
\ee
The remaining ingredient in $D$ is the function $E(t_1,t_2,t'')=\dq
R_{\qv}(t_1,t'')C_{\qv}(t_2,t'')$. This vanishes for $t''>t_1$ because
of the causality of $R_\qv$, and has a jump of size $\dq
C_\qv(t_2,t_1)=C(t_2,t_1)$ as $t''$ decreases past $t_1$. If $t_2<t_1$, $E$
actually remains constant at this value down to $t''=t_2$ because the
$t''$-dependence in $R_\qv(t_1,t'')C_\qv(t'',t_2)$ cancels as a
consequence of \eq{Rq} and\eq{C_twotime}. For $t''<\min(t_1,t_2)$, one
can use the same identities to express $E$ in terms of the kernel $K$:
the $\qv$-dependence (via $\omega$) of $R_\qv(t_1,t'')R_\qv(t_2,t'')$
is the same as that of $R_\qv^2((t_1+t_2)/2,t'')$, and accounting for
the remaining proportionality factors results in
$E(t_1,t_2,t'')=g((t_1+t_2)/2)g^{-1/2}(t_1)g^{-1/2}(t_2) K((t_1+t_2)/2,t'')$.
Carrying out the $t''$-integral in\eq{Dtilde} and exploiting the
decomposition\eq{L_pieces} of $L$ then gives for $D$ the general form
\bea
\fl D(t,\tw,t')&=&\half C(t,\tw)[\delta(t-t')+\delta(\tw-t')]+D_1(t,\tw,t')\theta(t'-\tw)\nn
\fl &&{}+D_2(t,\tw,t')\theta(\tw-t')
\label{D_pieces}
\eea
where the continuous pieces for $t'$ above and below $\tw$
respectively are
\be
\fl D_1(t,\tw,t')=\half C(t,\tw)\left(2T-
\int_{t'}^t dt''\, \Ltwo(t'',t')\right)
\label{D1}
\ee
and, abbreviating $\tbar=(t+\tw)/2$,
\bea
\fl D_2(t,\tw,t')&=& -\half C(t,\tw) \int_{\tw}^t dt''\,\Ltwo(t'',t')
+\frac{g(\tbar)}{\sqrt{g(t)g(\tw)}}\Biggl\{
\left(-\frac{\partial}{\partial t'} + 2T\right) K(\tbar,t')
\nn
\fl & &{}-\int_{t'}^{\tw} dt'' K(\tbar,t'')\Ltwo(t'',t')\Biggr\}
\label{D2_almost}
\\
\fl &=& -\half C(t,\tw) \int_{\tw}^t dt''\,\Ltwo(t'',t')
+ \frac{g(\tbar)}{\sqrt{g(t)g(\tw)}}
\int_{\tw}^{\tbar} dt'' K(\tbar,t'')\Ltwo(t'',t')
\label{D2}
\eea
The last simplification for $D_2$ arises because,
from\eq{L_definition} and\eq{L_pieces}, the terms in curly brackets
in\eq{D2_almost} would cancel exactly if the upper integration limit
was $\tbar$.

For the corresponding function $D^\chi$ for the susceptibility, the
$q$-integral in\eq{Dchi} can also be simplified by exploiting the
link\eq{integrated_chi} between $\chi_\qv$ and $R_\qv$:
\be
\dq R_{\qv}(t,t'') \chi_{\qv}(t'',\tw)=\chi(t,\tw)-\chi(t,t'')
\ee
This holds for $\tw<t''<t$; otherwise the function on the LHS vanishes
due to causality. Inserting into\eq{Dchi} and using again\eq{L_pieces}
gives
\bea
\fl D^{\chi}(t,\tw,t')&=&T\chi(t,\tw)\delta(t-t') + 
D_1^{\chi}(t,\tw,t')\theta(t'-\tw)+D_2^{\chi}(t,\tw,t')\theta(\tw-t')
\label{Dchi_pieces}
\eea
with
\bea
T^{-1}D_1^{\chi}(t,\tw,t')&=& -R(t,t') +
2T[\chi(t,\tw)-\chi(t,t')]
\nn
& &{}-\int_{t'}^t dt''\,\Ltwo(t'',t')[\chi(t,\tw)-\chi(t,t'')]
\label{D1chi}
\eea
and
\be
T^{-1}D_2^{\chi}(t,\tw,t')=
-\int_{\tw}^t dt''\,\Ltwo(t'',t')[\chi(t,\tw)-\chi(t,t'')]
\label{D2chi}
\ee
Note that the expressions above are general and valid for arbitrary
quenches, since we have not imposed any restrictions on the form of
response, correlation or the kernel $L$. They will therefore form the
basis for all further analysis of the correlation and susceptibility
variances $V_C$ and $V_\chi$ and their covariance $V_{C\chi}$.

In addition to the variances and covariance themselves we will also
consider the correlation coefficient
\be
\gamma=\frac{V_{C\chi}}{\sqrt{V_C V_{\chi}}}
\label{corr_coef}
\ee
which lies in the range $-1\ldots 1$; the extreme values correspond to
susceptibility and correlation fluctuations being fully correlated,
\ie\ identical up to a scale factor. The 
joint probability distribution of $(\hat{C},\hat{\chi})$ can be more
fully characterized by its contour lines. Due to the Gaussian nature
of the distribution (in our leading order approximation in $1/N$) the
contours are ellipes given by
\bea
N\left(
\begin{array}{ll}
\delta C & \delta \chi
\end{array}
\right)
\left(
\begin{array}{ll}
V_C & V_{C\chi}\\
V_{C\chi} & V_{\chi}
\end{array}
\right)^{-1}
\left(
\begin{array}{l}
\delta C\\
\delta \chi
\end{array}
\right)={\rm const}
\label{contours}
\eea 
These are centred on $(\delta C,\delta\chi)=(0,0)$, \ie\ on the mean
values $(C,\chi)$. Geometrically, it is then natural to define the
direction of the dominant fluctuations as the principal axis of
the ellipse. We define the negative slope of this as $X_{\rm fl}$, in
analogy with the FDR $X$ which gives the negative slope
of the FD plot relating the mean values $T\chi$ and $C$. If the
predictions for spin glasses summarized in the introduction also apply
to coarsening systems, one would expect $X_{\rm fl}$ to be close to $X$.
Explicitly, by diagonalizing the covariance matrix in\eq{contours} and
finding its largest eigenvector one has
\be
X_{\rm fl}=\frac{1}{2\gamma}
\left(\sqrt{\frac{V_C}{V_{\chi}}}-\sqrt{\frac{V_{\chi}}{V_C}}-
\sqrt{\left(\sqrt{\frac{V_C}{V_{\chi}}}-\sqrt{\frac{V_{\chi}}{V_C}}\right)^2+4\gamma^2}\right)
\label{X_fl}
\ee
In accordance with the definition of $X_{\rm fl}$ as the {\em
negative} slope of the principal axis,
it always has the opposite sign of the correlation coefficient
$\gamma$. We note that the definition of $X_{\rm fl}$, unlike that of
$\gamma$, depends in principle on the relative scaling of the axes of the FD
plot. The factors of $T$ included in\eq{variance_R}
and\eq{variance_cross} correspond to measuring the fluctuation slope
from contours in the $(\hat C,T\hat\chi)$ plane where the equilibrium
FDT is a line of slope $-1$. While not unique, this is certainly the
most natural choice.

\section{Quenches to $T>T_{\rm c}$}
\label{sec:hetero_above_Tc}

In this section we study quenches to above criticality so that, as 
discussed above, equilibrium is considered.
In equilibrium the average correlation and susceptibility functions
are time translation invariant (TTI) and 
related by the fluctuation-dissipation theorem.
We can ask whether FDT-like relations also hold for the fluctuations of 
correlation and susceptibility around their typical values, \ie\
whether $X_{\rm fl}$ is close to unity.

\subsection{Equilibrium expressions for $D$ and $D^{\chi}$}

In equilibrium, all functions depend only on time differences, so we
will write $K(t,\tw)=K(\dt)$, $L(t,\tw)=L(\dt)$ and so on, with
$\dt=t-\tw$. For the three-time functions we will keep the three
separate arguments; for $CC(t,\tw,t')$ this helps to 
avoid confusion with the two-time function
$CC(t,\tw)=CC(\dt)$.

In order to work out the equilibrium expressions for $D$ and
$D^{\chi}$, we need first the various covariance and response
functions, as well as the kernel $K$ and its inverse $L$. The Lagrange
multiplier $z$ approaches a constant value $z\eql$ at equilibrium,
corresponding to exponential growth $g(t)\propto\exp(2z\eql t)$ of the
function\eq{g_def}. One can then show that
$C_\qv(t,t)=T/(z\eql+\omega)$; since the spherical constraint imposes
$\dq C_{\qv}(t,t)=1$ at all times, $z\eql$ can be found from the condition
\be
\dq\frac{T}{z_{\rm eq}+\omega}=1.
\label{spherical_constraint}
\ee
For the moment we will leave the Lagrange multiplier unrestricted, so
that the following results will be valid for equilibrium at any
temperature $\geq \Tc$. (For $T<\Tc$ one would need to account separately
for the $\qv=\zv$ mode which acquires a nonzero expectation value
proportional to the equilibrium magnetization.) Later we will consider
first high temperatures, then generic temperatures above criticality,
and finally, in the next section, $T=\Tc$ where $z\eql$ vanishes.

The exponential behaviour of $g(t)$ in equilibrium reduces the Fourier
mode response and correlation functions\eq{Rq} and\eq{C_twotime} to the
simple forms
\bea
R_{\qv}(\dt)&=&e^{-(\om+z_{\rm eq})\dt}
\label{Req}\\
C_{\qv}(\dt)&=&\frac{T}{\om+z_{\rm eq}}e^{-(\om+z_{\rm eq})\dt}
\label{Ceq}
\eea
These determine the equilibrium form of the kernel $K$\eq{Kq_definition} as
\be
K(\dt)=\dq \frac{T}{\om+z\eql}e^{-2(\om+z\eql)\dt}  
\label{Keq}
\ee
and the (average) local correlation and response can be expressed in terms of
this as
\be
C(\dt)=\dq \frac{T}{\om+z\eql}e^{-(\om+z\eql)\dt} =
K\!\left(\textstyle{\frac{\dt}{2}}\right) 
\label{C_local_eq}
\ee
\be
TR(\dt)=T\dq e^{-(\om+z\eql)\dt} = -\half
K'\!\left(\textstyle{\frac{\dt}{2}}\right)
\label{R_local_eq}
\ee
\be
T\chi(\dt)=\dq
\frac{T}{\om+z\eql}\left(1-e^{-(\om+z\eql)\dt}\right)
= 1-K\!\left(\textstyle{\frac{\dt}{2}}\right) 
\label{chieq}
\ee
while the two and three time versions\eq{CC2times} and\eq{CC3times} of
$CC$ become
\bea
\fl CC(\dt) &=&\dq \left(\frac{T}{\om+z\eql}\right)^2
e^{-2(\om+z\eql)\dt} 
\label{CCeq2times}
\\
\fl CC(t,\tw,t')&=&\theta(t'-\tw)\dq \left(\frac{T}{\om+z\eql}\right)^2
e^{-(\om+z\eql)\dt}
\nn
\fl &&{}+\theta(\tw-t')\dq \left(\frac{T}{\om+z\eql}\right)^2
e^{-2(\om+z\eql)(\tbar-t')}
\\
\fl &=&CC(\dt/2)\theta(t'-\tw) + CC(\tbar-t')\theta(\tw-t')
\label{CCeq3times}
\eea
Notice that $CC(t,\tw,t')$ is independent of $t'$ in the regime $\tw<t'(<t)$.

Finally we need the inverse kernel $L$. Combining\eq{L_definition}
and\eq{L_pieces}, we can express its Laplace transform
$\hat\Ltwo(s)$ as
\be
\hat\Ltwo\eql(s)=s+2T-\frac{1}{\hat{K}\eql(s)}
\label{KL_Laplace}
\ee
We will require occasionally the integral of $L^{(2)}$ over all times,
which follows as
\be
\hat{L}^{(2)}(0)=\int_0^{\infty}dt\,L^{(2)}(t)=\left\{
\begin{array}{ll}
2T-\hat{K}^{-1}(0) & (T>\Tc \mbox{\ or\ } d>4)
\\
2T & (T=\Tc \mbox{\ and\ } d<4)
\end{array}
\right.
\label{Ltwo_0}
\ee
because $\hat{K}(0)=\int_0^{\infty} dt\,K(t)$ diverges at $T=\Tc$ for
$d<4$ (see Eq.\eq{K_power} below).
%
%

Putting everything together, we get from\eq{D_pieces},\eq{D1}
and\eq{D2} the explicit equilibrium form for $D$:
\bea
\fl D(t,\tw,t')&=&\half K\!\left(\textstyle\frac{\dt}{2}\right)[\delta(t-t')+\delta(\tw-t')]+D_1(t,\tw,t')\theta(t'-\tw)\nn
\fl &&{}+D_2(t,\tw,t')\theta(\tw-t')
\label{D_pieces_eq}
\eea
where
\be
\fl D_1(t,\tw,t')=\half K\!\left(\textstyle\frac{\dt}{2}\right)\left(2T-\int_0^{t-t'}d\tau\, \Ltwo(\tau)\right)
\label{D1_eq}
\ee
and
\be
\fl D_2(t,\tw,t')=-\half K\!\left(\textstyle\frac{\dt}{2}\right)\int_{\tw-t'}^{t-t'}d\tau\, L^{(2)}(\tau)
+\int_{\tw-t'}^{\tbar-t'}d\tau\, K\!\left(\tbar-t'-\tau\right)L^{(2)}(\tau)
\label{D2_eq}
\ee
Similarly one has from\eq{Dchi_pieces},\eq{D1chi} and\eq{D2chi} for
$D^{\chi}$
\bea
\fl D^{\chi}(t,\tw,t')&=&
\left[1\!-\!K\!\left(\textstyle{\frac{\dt}{2}}\right)\right]\delta(t-t') \!+\! 
D_1^{\chi}(t,\tw,t')\theta(t'-\tw)\!+\!D_2^{\chi}(t,\tw,t')\theta(\tw-t')
\label{Dchi_pieces_eq}
\eea
with
\bea
\fl D_1^{\chi}(t,\tw,t')&=&
\frac{1}{2}K'\!\left(\textstyle\frac{t-t'}{2}\right)+2T\left[K\!\left(\textstyle\frac{t-t'}{2}\right)-K\!\left(\textstyle\frac{\dt}{2}\right)\right]\nn
\fl &&{}-\int_{t'}^t dt''\,L^{(2)}(t''-t')
\left[K\!\left(\textstyle\frac{t-t''}{2}\right)-K\!\left(\textstyle\frac{\dt}{2}\right)\right]
\label{D1chi_eq}
\eea
and
\be
\fl D_2^{\chi}(t,\tw,t')=-\int_{\tw}^t dt''\,L^{(2)}(t''-t')
\left[K\!\left(\textstyle\frac{t-t''}{2}\right)-K\!\left(\textstyle\frac{\dt}{2}\right)\right]
\label{D2chi_eq}
\ee

\subsection{High $T$}
\label{sec:high_T}

Having derived the general equilibrium expression for the functions $D$ and 
$D^{\chi}$ at arbitrary temperature, we next study their time dependence
in the regime of temperatures above criticality, $T>\Tc$. First we
consider briefly the limit of high temperatures, where explicit
expressions can be obtained.

\begin{figure}
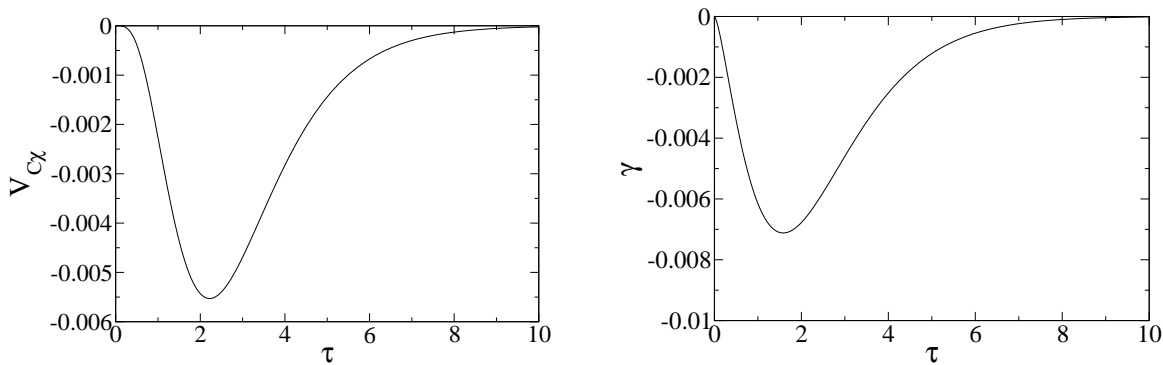

\setlength{\unitlength}{0.40mm} 
\begin{picture}(200,155)(-100,10)
\put(-88,10){\includegraphics[width=180\unitlength]{pVcross015.eps}}
\put(115,10){\includegraphics[width=180\unitlength]{pgammaHT.eps}}
\end{picture}
\caption{Plot of the covariance and the 
correlation coefficient $\gamma$ versus the rescaled time 
$\tau$ for $T=15$. 
These show a power law increase for small $\tau$ and an exponential 
decay for large $\tau$. Both are negative throughout, 
and the modulus of the correlation coefficient is 
much less than one, indicating that the fluctuations of correlation and
response functions are only weakly correlated.
}
\label{fig:crossHT}
\end{figure}

For $T\to\infty$, one sees from\eq{spherical_constraint} that the
Lagrange multiplier needs to scale as 
$z\eql=T+\order{(1)}$ because the frequencies $\omega$ are of order
unity and independent of $T$. This suggests a series expansion as
$z\eql=T+a+b/T+\order(1/T^2)$, and by substituting
into\eq{spherical_constraint} and using $\dq \omega = 2d$ and $\dq
\omega^2 = (2d)^2+2d$ one finds $-a=b=2d$.
The time dependence in the equilibrium 
functions\eq{Keq},\eq{CCeq2times} and\eq{CCeq3times} through the
combination $(z_{\rm eq}+\omega)\Delta t$ is then equal to $T\dt$ to
leading order. We therefore rescale the time difference with temperature
as $\tau=T\Delta t$ 
and expand all exponentials
$\exp[-(z\eql+\omega)\dt]=\exp[-(1+(\omega-2d)/T+2d/T^2)\tau]$ in
$1/T$. One finds in this way 
$K(\tau)=e^{-2\tau}(1+4d\tau^2/T^2)$, 
to $\order(1/T^2)$. The $\order(1/T^2)$ term in $K(\tau)$ is 
needed to determine the Laplace transform of $L^{(2)}$ 
from\eq{KL_Laplace}, because the leading order cancels. 
Inserting $K$ into\eq{KL_Laplace} shows that the first non vanishing 
term in $\hat{L}^{(2)}(s)$ is $\order(1/T)$, 
which transformed back to rescaled time variables 
yields the $\order(1)$ result%
\footnote{We remark that although $\order(1/T^2)$ terms are needed
to determine  
$L^{(2)}$, the latter has a value of $\order(1)$ and not
$\order(1/T^2)$ as we had mistakenly stated
in~\cite{AnnSol06}. Fortunately, this error had 
no effect on the calculations in~\cite{AnnSol06}, since the large-$T$
behaviour of $L^{(2)}$ was never used explicitly.}
$\Ltwo(\tau)=8d\exp(-2\tau)$. To use these results in a systematic
high-$T$ expansion up to $\order(1/T^2)$ of the correlation and susceptibility
(co-)variances, we need to know to which order in $1/T$ the functions
that appear need to be expanded. For large $T$ it is convenient to
rescale $D$ and  
$D^{\chi}$ by a factor $T$ in order to work with quantities of order unity.
The compensating factor $1/T$ is absorbed into the rescaling of the
time
integrations that lead from $D$ and $D^{\chi}$ to the (co-)variances. 
One can check that the terms proportional to $L^{(2)}$ in the rescaled 
$D$ and $D^{\chi}$ are smaller than the others by a factor $1/T^2$, because they are always 
obtained by integrating over time. Therefore 
$\Ltwo(\tau)$ is only needed to $\order(1)$.  
Expanding all other functions to order $\order(1/T^2)$ 
and inserting into\eq{VC_D},\eq{VR_D} and\eq{Vcross_D},
all integrals can be done explicitly. One obtains, 
after some lengthy but straightforward algebra, 
\be
V_C(\tau)=1-(1+2\tau)e^{-2\tau}
+\frac{1}{T^2}\left[2de^{-4\tau}-4de^{-2\tau}(1+2\tau^2+\tau^3)+2d\right]
\label{VC_HT}
\ee
\bea
V_{\chi}(\tau)&=&1+(3+2\tau)e^{-2\tau}-4e^{-\tau}+\frac{1}{T^2}
\left[\frac{2}{3}de^{-4\tau}-\frac{4}{3}de^{-3\tau}\right.
\nn
&&{}+{}\left.de^{-2\tau}(13+10\tau+6\tau^2+4\tau^3)-\frac{4}{3}de^{-\tau}(11+3\tau^2)+\frac{7}{3}d\right]
\label{VR_HT}
\eea
\bea
V_{C\chi}(\tau)&=&-\frac{2d}{3T^2}\bigl[2e^{-4\tau}-3e^{-3\tau}+6(1-\tau^2)
e^{-2\tau}+(6\tau-5)e^{-\tau}
\bigr]
\label{highTexpansion}
\eea
%
%
%
\begin{figure}
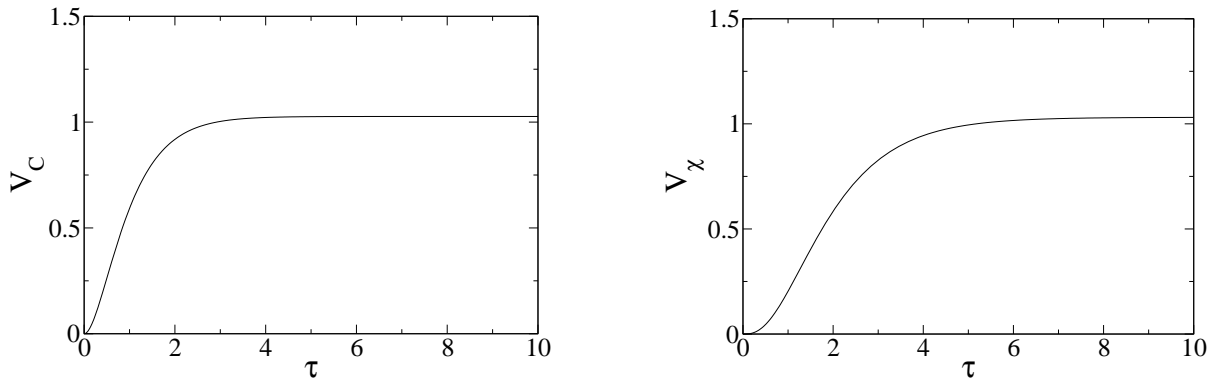

\setlength{\unitlength}{0.40mm} 
\begin{picture}(200,155)(-100,10)
\put(-103,10){\includegraphics[width=180\unitlength]{pVcc015.eps}}
\put(115,10){\includegraphics[width=180\unitlength]{pVchic015.eps}}
\end{picture}
\caption{Correlation and response variance versus the rescaled time $\tau$ 
for $T=15$. Both show a power law increase for small $\tau$ and an
exponential approach to their limit value for large $\tau$.}
\label{fig:CRHT}
\end{figure}
Plots of\eq{VC_HT},\eq{VR_HT} and\eq{highTexpansion} 
are shown in figures~\ref{fig:CRHT} and~\ref{fig:crossHT} (left) for 
$T=15$. 
One can study the high temperature limit of the correlation and 
susceptibility variances directly by setting 
the $\order{(1/T^2)}$ corrections in\eq{VC_HT} and\eq{VR_HT} to zero. 
This shows that for high $T$ the correlation
and susceptibility variances are monotonically increasing funtions of
$\tau$, starting from zero at $\tau=0$ ($\ie$ $\dt=0$). This is as
expected since $\hat C(\tw,\tw)=1$ cannot fluctuate due to the
spherical constraint, while $\hat\chi(\tw,\tw)$ vanishes trivially. An
expansion for small $\tau$ shows that the
correlation and susceptibility variances increase initially as,
respectively, $V_C(\tau)=2\tau^2$ and $V_{\chi}(\tau)=(2/3)\tau^3$.
These scalings, including the prefactors, will also be found at finite
temperature (see below). They show
that there are significant correlations in the time evolution of $\hat
C$ and $\hat\chi$; if the fluctuations had independent increments at
different times this would lead to a random walk for the fluctuations
and hence a much more rapid increase of the variances, $V\sim \tau$.

We cannot infer from the $1/T$ expansion whether the monotonic
behaviour in $\tau$ of the variances holds also for finite
temperature. However, the fact that already the leading $1/T^2$ corrections
are non-monotonic in $\tau$ suggests that the overall variation at
finite $T$ may also be non-monotonic.
Indeed, for the correlation variance $V_C$ we will see in
Sec.~\ref{sec:equilibrium_d_gt_4} using different arguments that a
non-monotonic dependence on $\dt$ (or equivalently $\tau$) occurs at
least in $d>4$ and for $T$ not too far above $\Tc$.

In the limit $T\to\infty$, 
both variances approach the constant
value $1$ exponentially fast in $\tau$.
For the correlation this can be
explained relatively simply: as the spins $S_i(\tw)$ and $S_i(t)$
decorrelate at long times and are also uncorrelated in space for
large $T$, $\hat C = (1/N)\sum_i S_i(t)S_i(\tw)$ becomes a zero mean
Gaussian random variable of variance $1/N$. Consistent with this
intuition, the dominant contribution to $V_C$ for large $\dt$ comes
from the Gaussian fluctuations which are described by the first two
terms in\eq{VC_D}; in fact, only the first term survives for
$\dt\to\infty$. It should be emphasized, however, that the high-$T$
limit does not amount to neglecting all non-Gaussian effects. Indeed,
the Gaussian terms from\eq{VC_D} would give the quite incorrect result
$V_C=2$ for $\dt=0$.

We next look at the covariance of correlation and susceptibility, and
the consequences for the correlation coefficient $\gamma$ and the
fluctuation FDR $X_{\rm fl}$.  
Eq.\eq{highTexpansion} shows that the covariance is $\order{(1/T^2)}$ 
for any finite $\tau$, and it vanishes in the limits of both small and
large $\tau$ as, respectively,
\be 
V_{C\chi}(\tau)\approx -4d\tau^4/(3T^2)
\label{CrossShortTau}
\ee
and $V_{C\chi}\approx -4d\tau e^{-\tau}/T^2$. 
Plotting the full expression\eq{highTexpansion} (see
Fig.~\ref{fig:crossHT} left) shows that $V_{C\chi}$ is 
negative not just in these two limits but in fact for all $\tau$.

From the above results one can determine the 
correlation coefficient, as defined in\eq{corr_coef}, for high $T$.
For $\tau\rightarrow 0$ and in the limit of high temperature, one obtains
directly from the small-$\tau$ scaling of the (co-)variances
that the correlation coefficient goes to zero as $\gamma\sim
-\tau^{3/2}/T^2$. 
For the opposite limit 
$\tau\rightarrow\infty$ of long times,
$V_C=V_{C\chi}=1$ to leading order, as explained. 
This yields $\gamma\approx
V_{C\chi}\approx -4d\tau e^{-\tau}/T^2$. 
A plot of $\gamma$ (see Fig.~\ref{fig:crossHT} right) 
shows that like $V_{C\chi}$ it is negative for all
$\tau$, and its modulus is smaller than unity as it should be. The
scaling with $1/T^2$ shows that, for high temperatures, correlation
and susceptibility fluctuations become increasingly less correlated
with each other.
%

Studying the fluctuation FDR\eq{X_fl} to characterize 
the joint distribution of correlation and susceptibility requires some
care. If we first proceed as above, keeping $\tau$ of order unity
fixed and taking $T\to\infty$, then $\gamma$ scales with $1/T^2$ as
we saw earlier and so becomes small compared to the other terms of order
unity in\eq{X_fl}. We can then expand in $\gamma$ to get to leading order
\be
X_{\rm fl}=-\frac{\gamma}
{\sqrt{{V_{C}}/{V_{\chi}}}-\sqrt{{V_{\chi}}/{V_C}}}
\label{Linear}
\ee 
For small $\tau$, where $\gamma\sim
-\tau^{3/2}/T^2$ and $V_{\chi}/V_C \sim \tau$, this gives
\be
X_{\rm fl}\sim \frac{\tau^2}{T^2}
\label{X_fl_highT}
\ee
For large $\tau$, the denominator of\eq{Linear}
goes to zero even faster than the numerator, and we find 
\be
X_{\rm fl} \sim \frac{\tau}{T^2}
\label{Linear2}
\ee
However, this result must clearly break down when $\tau$
becomes too large at finite $T$, as Eq.\eq{Linear} was predicated on
$\gamma$ being small compared to
$(V_C/V_\chi)^{1/2}-(V_\chi/V_C)^{1/2}$. To understand
what happens in this regime, we use that $V_C=1+2d/T^2
$ and $V_{C\chi}=1+7d/(3T^2)
$ to leading order for $\tau\to\infty$ at finite $T$, where we need to
keep the $\order(1/T^2)$ corrections. Then
$\gamma=-4d\tau e^{-\tau}/T^2$ in the outer square root 
of\eq{X_fl} can be neglected as smaller than the 
other term under this root, giving to leading order a
temperature-independent exponential increase
\be
X_{\rm fl}=\frac{\sqrt{{V_{C}}/{V_{\chi}}}-\sqrt{{V_{\chi}}/{V_C}}}
{\gamma}=\frac{e^{\tau}}{12\tau}
\label{Exponential}
\ee

\begin{figure}
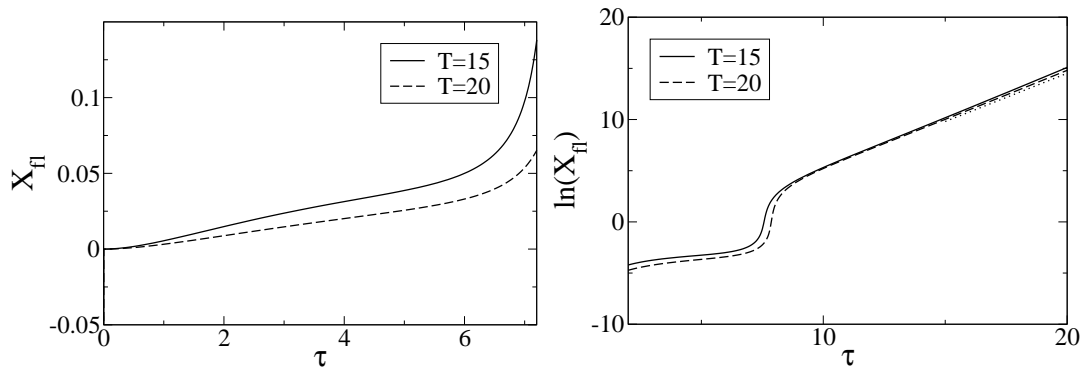

\centerline{\includegraphics[width=7.0cm,clip=true]{pXfllinPS.eps}\,
\includegraphics[width=7.0cm,clip=true]{pXfllogPS.eps}}
\caption{Left: Plot of the fluctuation 
FDR $X_{\rm fl}$ versus the rescaled time $\tau$, for 
$d=3$ and $T=15, 20$ as shown in the legend. $X_{\rm fl}$ starts off
quadratically  
for small $\tau$, then becomes linear, its slope increasing with
decreasing $T$ in both regimes. Eventually it crosses over into a regime of
exponential growth, which is not yet visible in the
$\tau$-range shown. Right: Plot of $\ln(X_{\rm fl})$ versus $\tau$, at
$T=15, 20$ as indicated in the legend. This shows the crossover to the
regime of $T$-independent exponential increase at large $\tau$, which
takes place at $\tau\approx 2\ln T$ and is represented by the dotted line 
on the right of the graph. 
\label{fig:FDHighT_Xfl}
}
\end{figure}

The crossover between the linear and exponential regimes, Eq.\eq{Linear2} 
and Eq.\eq{Exponential} respectively,
can be shown to be due to the competition, for large $T$ and $\tau$,
between $\order{(1/T^2)}$ and $\order{(e^{-\tau})}$ terms 
in\eq{VC_HT} and\eq{VR_HT}, and therefore in\eq{X_fl},
and takes place at $\tau\approx 2\ln T$. This is shown in 
Fig.~\ref{fig:FDHighT_Xfl} on the right, along with (on the left) the
crossover between the quadratic 
and the linear regimes, Eq.\eq{X_fl_highT} and Eq.\eq{Linear2}, that
occurs at shorter times.

\begin{figure}
\centerline{\includegraphics[width=7.0cm,clip=true]{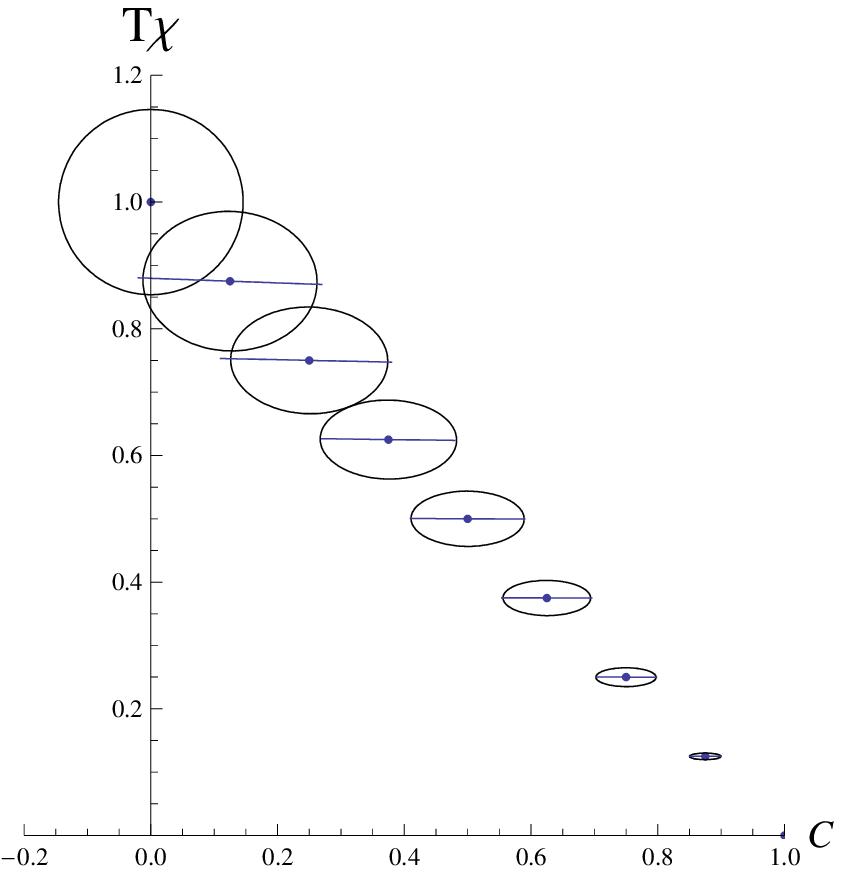}\,
\includegraphics[width=7.0cm,clip=true]{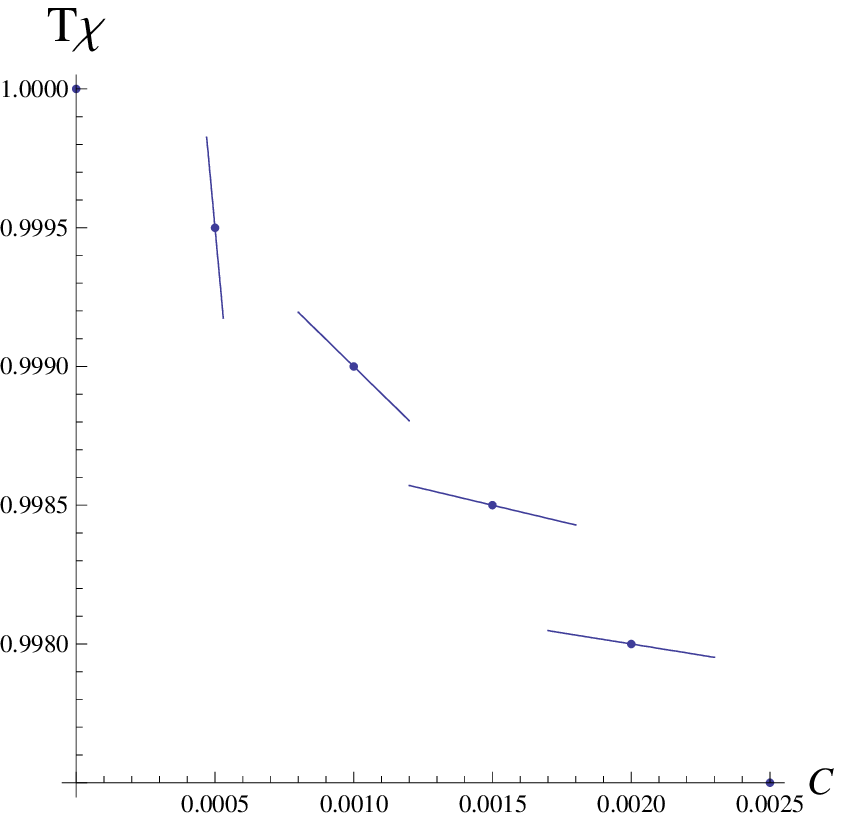}}
\caption{Left: Fluctuations of the coarse-grained local correlation $\hat C$
and susceptibility $T\hat\chi$ in the high temperature limit. The mean
values, indicated by the centres of the ellipses, lie on the straight
equilibrium FDT line. The ellipses themselves show contour lines for several
different mean values, corresponding to different scaled time
differences $\tau=T\dt$; we chose $d=3$, 
$T=10$, $N=50$ and set the constant on the 
RHS of\eq{contours} to unity. 
The principal axis of each ellipse is shown, and represents
the direction of the biggest fluctuations. This lies on the
correlation axis for small $\tau$ ($X_{\rm fl}\ll 1$) but rotates to
lie along the susceptibility axis ($X_{\rm fl}\gg 1$) for large 
$\tau$. This rotation of the principal axis is continuous but in this
graphical representation happens very quickly in the top left corner
of the plot: this is because it 
takes place at large values of $\tau$, i.e.\ when the mean correlation
has already  
decayed to a very small value. The fluctuation contours in this regime
are close to circles and would lie essentially on top of the leftmost
ellipse shown. Right: A zoom of the top left corner of the
plot on the left. Here only the direction of the principal axis of
each ellipse is drawn, to make the rotation towards the $\chi$-axis clearer.
\label{fig:FDHighT}
}
\end{figure}

With the expression of the (co-)variances to the required orders at hand, 
we now plot the contour lines of the joint distribution of the
fluctuating correlation and susceptibility $(\hat C, T\hat\chi)$. The
mean values $C=\langle\hat C\rangle$ and $\chi=\langle\hat\chi\rangle$
can be read off from\eq{C_local_eq} and\eq{chieq} as
$C(\tau)=\exp(-\tau)=1-T\chi(\tau)$ and produce the straight
line of slope $-1$ expected from equilibrium FDT. 
(Note that we do not need to normalize the FD plot because for
our local spin correlations the equal time correlator $C(t,t)=1$
always.) Fig.~\ref{fig:FDHighT} shows contour lines of the fluctuation
distributions for a range of different mean values (\ie\ different
$\tau$); their centres lie on the straight equilibrium FDT line.
For the purposes of this graphical illustration, we have aimed to
choose a relatively low temperature, as otherwise the covariance
becomes too small and all fluctuation contours degenerate into
ellipses oriented along the $C$ and $\chi$-axes. We cannot go too low,
of course, as otherwise truncating the $1/T$ expansion cannot be
justified. The choice $T=10$ is a reasonable compromise: the
$\order(1/T^2)$ corrections to the variances are then significantly
smaller than the leading order terms. However, $\order(1/T^3)$ terms
-- which can be worked out along the lines above -- are comparable to
the $\order(1/T^2)$ contributions, so the results shown are not fully
quantitative.
We consider $d=3$ and set the scale of the contours by taking $N=50$
and unity for the constant on the RHS of\eq{contours}. The relatively
small value of $N$ was taken only for better visibility; a larger
value would simply shrink all ellipses uniformly.

In summary, the fluctuations of correlation and susceptibility are
not linked in a manner akin to the equilibrium FDT. The dominant fluctuation
direction measured by $X_{\rm fl}$ does not lie on the straight line 
of slope $-1$ that locates the average quantities. Instead, this
direction is along the horizontal (correlation) axis for small times
and along the vertical (susceptibility) axis for large times.

\subsection{Small time differences}
\label{sec:short_time_differences}

Next we consider the short-time behaviour of the correlation and
susceptibility variances at some generic temperature $T\geq \Tc$.
For the correlation one expects from the high-$T$ result that the
leading term for small $\dt$ will be $\order(\dt^2)$. We therefore
expand $D$ as 
\bea
\fl D &=&
\left[\half-\frac{T\dt}{2}+\frac{K''(0)}{16}\dt^2\right][\delta(t\!-\!t')+\delta(\tw\!-\!t')]
+ \theta(t'\!-\!\tw)D_1 + \theta(\tw\!-\!t')D_2
\\
\fl D_1&=&T-T^2\dt -
\half L^{(2)}(0)(t-t')
\\
\fl D_2&=&\frac{1}{8}[2TL^{(2)}(\tw-t')-L^{(2)}{}'(\tw-t')]
\dt^2
\eea
Here we have used\eq{properties_K}; 
$D_1$ needs to be expanded only to
linear order in quantities of $\order(\dt)$ (including $t-t'$) because
it is integrated over the range $\tw<t'<t$ which is itself of $\order(\dt)$.
Inserting into\eq{VC_D} and expanding
the remaining functions\eq{CCeq2times} and\eq{CCeq3times},
the various terms involving $L^{(2)}$ cancel to leading order.
In fact, to $\order{(\Delta t^2)}$, $D_2$ 
only contributes 
to the single integral 
in\eq{VC_D}, where the remaining three-time function $CC$ is 
set to its zero-order value, 
and to the double integral when it is combined with the $\delta$-terms 
appearing in the definition of $D$;
however, these contributions cancel. The remaining quantities depending on 
$L^{(2)}$ are found in $D_1$ and are already $\order{(\Delta t)}$, 
so they provide contributions of $\order{(\Delta t^2)}$ when integrated 
over $\tw<t'<t$, if the remaining quantities in the integral are 
$\order{(1)}$. 
Again, this can only be realized in the single 
integral by setting the three time function $CC$ 
to its zero order value, and in the coupling with the 
$\delta$-terms in the double integral, but these terms cancel. 
Also, in the final result there is 
a leading order cancellation of the terms proportional to $K''(0)$ and one obtains
\be
V_C(\dt)=2T^2\Delta t^2
\label{CshortTime}
\ee
This agrees in both the $\dt$-dependence and the prefactor with the
result of the high-$T$ expansion. It is worth noting that in order to get the 
short time behaviour of the correlation variance, we have expanded 
$CC(t'-\tw')$ around $CC(0)$. This quantity diverges in $d<4$ at
$T=\Tc$. But the fact that it cancels from the leading short-time
behaviour should mean that\eq{CshortTime} remains valid: in principle
we just need to regularize in some way, e.g.\ by keeping $\tw$ large
but finite, then perform the short-time expansion and finally remove the
regularization, which should lead back to\eq{CshortTime}.

For the susceptibility variance, as given in\eq{VR_D}, 
we can expand $D^\chi$ as
\bea
\fl D^{\chi} =
\dt\,\delta(t-t')-\theta(t'-\tw)+\order(\dt)\theta(t'-\tw)+\order(\dt^2)\theta(\tw-t')
\label{Dchi_small_dt}
\eea
The first two terms and the remainder give contributions of
$\order(\dt)$ and $\order(\dt^2)$ respectively to the integral over $t'$. Keeping
only the first two terms and replacing $CC(t'-\tw')$ by $CC(0)$
in\eq{VR_D} should then give the leading term in $V_\chi$ of
$\order(\dt^2)$; but this cancels because $\int dt'
[\dt\,\delta(t-t')-\theta(t'-\tw)] = 0$. The same argument shows that to
$\order(\dt^3)$ the cross terms between the $\order(\dt)$ and
$\order(\dt^2)$ contributions from\eq{Dchi_small_dt} cancel. The only
remaining $\order(\dt^3)$ term is then
\bea
\fl V_\chi &=& \half\int_{\tw}^t dt'd\tw'[\dt\,\delta(t-t')-\theta(t'-\tw)]
[\dt\,\delta(t-\tw')-\theta(\tw'-\tw)]CC(t'-\tw')
\\
\fl &=& \half \dt^2 CC(0) - \dt\int_{\tw}^t dt'\,CC(t-t') + 
\int_{\tw}^t dt'\int_{\tw}^{t'}d\tw'\,CC(t'-\tw')
\eea
One now expands $CC(t'-\tw')=CC(0)-2T(t'-\tw')$ for $t'\geq \tw'$,
using\eq{CCeq2times}, to find that the $\dt^2CC(0)$-terms cancel as
expected, leaving
\be
V_{\chi}(\dt)=\frac{2}{3}T^3\Delta t^3
\label{RshortTime}
\ee
This again agrees with the high-$T$ expansion in both the scaling with
$\dt$ and the prefactor.

For the covariance,
finally, performing the perturbation expansion in small $\dt$ shows that 
the leading short-time contribution is only of fourth order in $\dt$. 
The temperature dependence of the prefactor is quite complicated and
involves the kernel $L^{(2)}$. We only show here the limit of the
prefactor for large $T$, which is 
\be 
V_{C\chi}(\dt)=-\frac{4dT^2}{3}\dt^4
\label{CrossShortTime}
\ee
in agreement with\eq{CrossShortTau}.
Correspondingly, the correlation coefficient is again negative,
growing in modulus initially as $\gamma\sim-\Delta t^{3/2}$. The
fluctuation slope $X_{\rm fl}$ can also be expanded as in the high-$T$
case\eq{X_fl_highT}, leading again to a quadratic short-time increase
\be
X_{\rm fl}\sim -\frac{V_{C\chi}}{V_C} \sim \dt^2
\ee

\subsection{Large time differences}

We next turn to the behaviour of the correlation and susceptibility
fluctuations at large time differences, at equilibrium at temperatures
{\em above} criticality; the dynamics at criticality exhibits
qualitative differences and is considered separately in the next section.
The asymptotic behaviour of (the equilibrium forms of) $K$, $\Ltwo$ and $CC$
is an exponential decay. For $K$ and $CC$ this follows directly
from\eq{Keq} and\eq{CCeq2times}, where all Fourier modes decay as
$\exp(-2z\eql\dt)$ or faster. The Laplace transform $\hat K(s)$ then
has all its singularities bounded away from $s=0$, and the same
follows for $\hat\Ltwo(s)$ from\eq{KL_Laplace}.
Thus, looking at $D(t,\tw,t')$ as given by\eq{D_pieces_eq},
the $K(\dt/2)$-prefactor ensures that the $\delta$-contributions to $D$ decay
exponentially for large time differences, and the same is true for the
$D_1$-term found in\eq{D1_eq}. The $D_2$-term given in\eq{D2_eq} has 
the same behaviour. This is obvious for the first term; for the second
term, bounding both $K(t)$ and $\Ltwo(t)$ by $\exp(-ct)$ shows that
the integral is bounded by $\dt\exp[-c(\tbar-t')]\leq\dt\exp(-c\dt/2)$.
Thus, for large $\dt$, all the non-Gaussian corrections in\eq{VC_D},
as well as the two-time Gaussian term $CC(t,\tw)$, 
decay exponentially to zero; the asymptotic value 
of the correlation variance is then given by the
time-independent Gaussian term 
$\dq C_{\qv}(t,t)C_{\qv}(\tw,\tw)=\dq T^2/(\om+z_{\rm eq})^2$.
This increases as the temperature is reduced 
towards $T_{\rm c}$; the limit value for $T\to\Tc$ is finite for $d>4$
but infinite 
for $d<4$. From the reasoning above it follows that for $T>\Tc$ the approach
of the correlation variance to its
asymptotic value for $\dt\to\infty$ is exponential in $\dt$, up to
power law factors.

Analogous reasoning for $D^{\chi}$ in\eq{Dchi_pieces_eq} leads one to
discard as subleading for large $\dt$ the $D_2^{\chi}$ term in\eq{D2chi_eq}
and the terms proportional to $K(\dt/2)$ appearing in\eq{D1chi_eq} for 
$D_1^{\chi}$. The
asymptotic susceptibility variance can therefore be found from\eq{VR_D} by 
replacing $D^{\chi}$ with
\be
\fl D_{\rm
short}^{\chi}=\delta(t-t')+\half K'\!\left(\textstyle\frac{t-t'}{2}\right)
+2TK\!\left(\textstyle\frac{t-t'}{2}\right)
-\int_{t'}^t dt''\,L^{(2)}(t''-t')
K\!\left(\textstyle\frac{t-t''}{2}\right)
\label{Dshort}
\ee 
We have added the subscript ``short'' because all contributions
retained in $D_{\rm short}^\chi$ decay exponentially as $t-t'$ increases,
concentrating the ``mass'' of the 
integrals over $t'$ and $\tw'$ in\eq{VR_D} into the regions $t-t'=\order{(1)}$ 
and $t-\tw'=\order{(1)}$. (The factor $CC(t'-\tw')$ does not affect
this reasoning as its values are also largest when $t'$ is close to
$\tw'$.) Nevertheless, because $CC(t'-\tw')$ does vary significantly on
$\order(1)$ timescales one cannot simplify the expression for the
asymptotic value of $V_\chi$ further, beyond the replacement of
$D^\chi$ by $D^\chi_{\rm short}$ in\eq{VR_D}. Barring accidental
cancellations (which, by continuity with the nonzero result for
$T\to\infty$, one does not expect), the result will be nonzero. The
approach to the limit will again be exponential in $\dt$.

Finally, in the cross-correlation\eq{Vcross_D} there are no
$\order{(1)}$ contributions that survive in the long-time limit,
because every term is proportional to $D$ and thus decays
exponentially with $\dt$. Qualitatively, then, the behaviour at finite
$T>\Tc$ is the same as that for $T\to\infty$, with the variances of
correlation and susceptibility approaching nonzero asymptotic values
exponentially fast in $\dt$, and the covariance decaying to zero (from
below) in the same manner.
 
\section{Quenches to $T_{\rm c}$, $d>4$}
\label{sec:hetero_dgt4}

Above we derived the large-$N$ statistics of the correlation and susceptibility fluctuations for quenches to 
temperatures above criticality. In this section and in Sec.~\ref{sec:d_lt_4} 
we consider 
quenches directly to criticality, so from now on $T=\Tc\equiv (\dq
\,1/\om)^{-1}$. In principle we then expect aging
effects~\cite{GodLuc00b,AnnSol06}, but it will turn out that for the
fluctuation statistics these are largely negligible as long as we are in
dimension $d>4$. We therefore consider
first the situation in equilibrium at criticality in $d>4$, focussing on large
time differences $\dt$; the short-time limit does not need to be
analysed again here because the results in
Sec.~\ref{sec:short_time_differences} apply even at $T=\Tc$. As discussed
in more detail in Sec.~\ref{sec:d_lt_4}, for $d<4$ one has to keep $\tw$
finite to avoid the appearance of infinite terms, \ie\ one has to look
directly at the non-equilibrium situation.

\subsection{Equilibrium}
\label{sec:equilibrium_d_gt_4}

The asymptotic behaviour of the equilibrium forms of $K$, $CC$ and $\Ltwo$
for $T\rightarrow T_{\rm c}$ is quite different from the high
temperature phase 
because $z_{\rm eq}$ vanishes. In the equilibrium form\eq{Keq} of the
kernel $K$, the integral is for
large time-differences dominated by small $\om$.
Because $\omega\approx q^2$
for small $q=|\qv|$, the phase space factor in the $\qv$-integrals is
$(dq)=\sigma_d d\omega\, \omega^{d/2-1}$ for small $q$ or $\omega$, 
with the proportionality constant
\be
\sigma_d=(4\pi)^{-d/2}\Gamma^{-1}(d/2)
\ee
Then from\eq{Kq_definition} one finds for large $\dt$
\be
K(\dt)=\sigma_d \int d\om \,\om^{(d-2)/2} \frac{T}{\om}
e^{-2\om\dt}  
=k_d \dt^{(2-d)/2}
\label{K_power}
\ee
and correspondingly, from the small-$s$ expansion of\eq{KL_Laplace},
\be
\Ltwo(\dt)=\lambda_d \left\{
\begin{array}{ll}
\dt^{(2-d)/2} & \mbox{for $d>4$} \\
\dt^{(d-6)/2} & \mbox{for $d<4$}
\end{array}
\right.
\label{L_power}
\ee
with $k_d$ and $\lambda_d$ $d$-dependent constants~\cite{AnnSol06}.

In equilibrium at criticality the function $CC(\dt)$
also decays as a power law for $\dt\gg1$. This can easily be worked
out in $d>4$:
\be
\fl CC(\dt) = \dq \frac{T^2}{\omega^2}e^{-2\omega \dt} = 2T
\int_{\dt}^\infty dt'\, K(t') \sim \dt^{(4-d)/2}
\label{CC_eql}
\ee
The three-time function $CC(t,\tw,t')$ given in\eq{CCeq3times}
increases with $t'$ up to $t'=\tw$; in the range $t'=\tw\ldots t$ it then
remains constant and equal to $CC(\dt/2)$.

For the correlation variance\eq{VC_D}, the first of the two Gaussian
terms is a constant of order unity, $\dq
C_{\qv}(t,t)C_{\qv}(\tw,\tw)=\dq T^2/\om^2$. The second one,
$CC(\dt)$, is proportional to $\dt^{(4-d)/2}$ for large $\dt$
from\eq{CC_eql}. One can
show that these are in fact the two leading terms, with the non-Gaussian
corrections contributing at most $\order(\dt^{4-d})$
asymptotically.
(See~\cite{AnnSol08} for a detailed discussion of the scaling of these 
terms.)
Note that because the prefactor of
$CC(\dt)\sim
\dt^{(4-d)/2}$ in\eq{VC_D} is {\em positive}, $V_C$ approaches its
(positive) asymptotic value from {\em above}; given that it starts at
zero at $\dt=0$, it is therefore non-monotonic in $\dt$. Because
$V_C(\dt)$ must depend continuously on temperature, at least at finite
$\dt$, this non-monotonicity must then be present also in a range of
temperatures above $\Tc$.

To understand the asymptotics of the susceptibility variance\eq{VR_D}
it is useful to decompose
\be
D^\chi(t,\tw,t') = D^\chi_{\rm short}(t,\tw,t') + D^\chi_{\rm long}(t,\tw,t')
\label{Dchi_short_long_decomposition}
\ee
where
\be
\fl D^\chi_{\rm long}(t,\tw,t') = -
K\!\left(\textstyle\frac{\dt}{2}\right)\delta(t-t') \!+\!
D^{\chi}_{1,\rm long}(t,\tw,t')\theta(t'-\tw)\!+\!D_2^\chi(t,\tw,t')\theta(\tw-t')
\label{Dchi_long}
\ee
and
\be
D^{\chi}_{1,\rm long}(t,\tw,t') =
-K\!\left(\textstyle\frac{\dt}{2}\right)\left[2T
-\int_{t'}^t dt''\,L^{(2)}(t''-t')\right]
\ee
while $D^\chi_2$ is as written in\eq{D2chi_eq}. The short-time part
of\eq{Dchi_short_long_decomposition}, defined in\eq{Dshort}, decays on
timescales $t-t'=\order(1)$ and its integral over $t'$ is of order
unity. More precisely, Laplace transforming $D^\chi_{\rm
short}(t,\tw,t')$ w.r.t.\ $t-t'$ and expanding for small $s$ gives
\bea
\fl \hat{D}_{\rm short}^{\chi}(s)&=&2\hat{K}(2s)[s+2T-\hat{L}(s)]=
2\frac{\hat{K}(2s)}{\hat{K}(s)}\approx 
2\frac{\hat{K}(0)-c(2s)^{(d-4)/2}}{\hat{K}(0)-cs^{(d-4)/2}}
\\
\fl&\approx &2+2\frac{c}{\hat{K}(0)}s^{(d-4)/2}(1-2^{(d-4)/2})
\label{D_short_s}
\eea
where we used\eq{KL_Laplace} and the small $s$-expansion of
$\hat{K}(s)=\hat{K}(0)-cs^{(d-4)/2}$~\cite{AnnSol06}. This shows that
the integral of $D^\chi_{\rm short}(t-t')$ over all $t-t'$ equals 2, and
that $D^\chi_{\rm short}(t-t')$ decays as $(t-t')^{(2-d)/2}$ for large $\dt$.
The complementary long-time part $D^\chi_{\rm long}$, on the other
hand, can be seen to have a 
structure similar to that of $D$: the $\delta$-term has weight $\sim
\dt^{(2-d)/2}$, $D^\chi_{1,\rm long}$ is of order $\dt^{(2-d)/2}$ (and
constant for $t-t'\gg 1$), and $D^\chi_2$ is of order $\Ltwo(t-t')\sim
(t-t')^{(2-d)/2}$. Decomposing now the product $D^\chi D^\chi$
in\eq{VR_D} according to\eq{Dchi_short_long_decomposition}, the
$D^\chi_{\rm short}D^\chi_{\rm short}$ term gives the leading
asymptotic term of $V_\chi$, a constant of order unity. 
We omit here the detailed analysis of the scaling of the other terms,
which can be found in~\cite{AnnSol08},
and point out only that overall $V_\chi$ approaches
its asymptotic value via leading power law terms $\sim\dt^{(2-d)/2}$
and $\sim\dt^{4-d}$; the former dominates for $d>6$, the latter for
$4<d<6$. There appears to be no simple way of estimating the sign of
these terms to verify whether $V_\chi$, like $V_C$, has a non-monotonic
$\dt$-dependence.

In the covariance\eq{Vcross_D} there are, as in the case of $T>\Tc$,  no
$\order{(1)}$ terms that
survive for long times. To obtain the leading decay
to zero, one considers first the single integral involving $D^\chi$ and the
three-time function $CC(t,\tw,t')$. 
The leading contribution arises from the
short-time part $D^\chi_{\rm short}$ of $D^\chi$, giving $\approx
-2CC(t,\tw,t)=-2CC(t-\tw)\sim -\dt^{(4-d)/2}$,
%
so that $V_{C\chi}(\dt)$ approaches $0$ from below as $-\dt^{(4-d)/2}$. 
The scaling of the remaining subleading terms is analysed in~\cite{AnnSol08}.

\begin{figure}
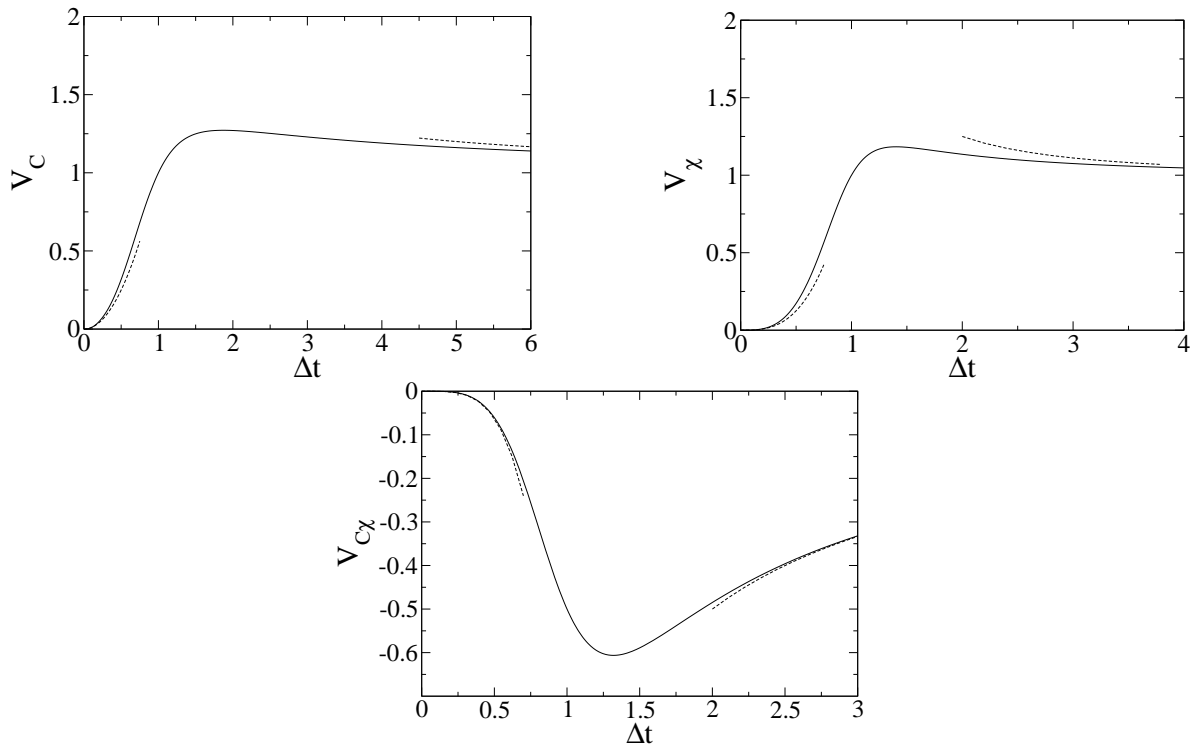

\includegraphics[width=7.0cm,clip=true]{0708c_above4.eps}
\includegraphics[width=7.0cm,clip=true]{0708r_above4.eps}
\centerline{\includegraphics[width=7.0cm,clip=true]{CX2_above4.eps}}
\caption{Schematic plot of the $\dt$-dependence of the 
variances, $V_C$ and $V_{\chi}$, and covariance, $V_{C\chi}$, 
for critical coarsening and $d>4$. These functions increase from $0$
to $\order{(1)}$ values as power laws of $\dt$ given 
respectively in\eq{CshortTime},\eq{RshortTime} and\eq{CrossShortTime}, 
and then decay asymptotically with negative powers of $\dt$. $V_C$ and
$V_{C\chi}$ are non-monotonic; for $V_\chi$ it is not obvious whether
the asymptotic value is approached from above or below.}
\label{fig:above4_Tc}
\end{figure}

A schematic plot of the full $\dt$-dependence of the variances and the 
covariance at criticality and $d>4$ is shown in Fig.~\ref{fig:above4_Tc}.
The non-monotonicity of the correlation variance (and, possibly, the
susceptibility variance) is consistent with a scenario in which 
for both short and long time regimes local correlations and responses
display minor fluctuations across the system and dynamical 
histories, whereas for intermediate 
time they depend strongly on the particular state reached by the system 
in a given time interval, making fluctuations across dynamical 
histories large. This type of behaviour is quite generic: indeed,
as explained after\eq{VC_chi4}, $V_C$ is the four-point function commonly used to
quantify dynamic heterogeneities, and this is found to exhibit
a maximum as a function of $\dt$ for many systems with slow
dynamics~\cite{GloJanLooMacPoo98,FraDonParGlo99,LacStaSchNovGlo02}. Note though that typically (\eg\ in
coarsening below $\Tc$~\cite{MayBisBerCipGarSolTra04,MaySolBerGar05}) the position of the
maximum scales with the age of the system whereas here, for coarsening
at criticality above the upper critical dimension, it occurs for an
{\em age-independent} time difference $\dt$.

Gathering the above results for $\dt\rightarrow \infty$, 
$V_C(\dt)\rightarrow\order{(1)}$, 
$V_{\chi}(\dt)\rightarrow\order{(1)}$ and 
$V_{C\chi}(\dt)\rightarrow-\dt^{(4-d)/2}$, 
yields for the correlation coefficient the power law decrease
$\gamma\sim-\dt^{(4-d)/2}$.
The FDR for the local fluctuations $X_{\rm fl}$ as given in\eq{X_fl}
is then positive, but to say more one would need to know whether $V_C$
and $V_\chi$ is bigger for $\dt\to\infty$. If $V_C>V_\chi$, the
elliptical contour of $P(\hat C, T\hat\chi)$ has its main axis along
the $\hat C$ direction and, as one can show formally by expanding\eq{X_fl} for
small $\gamma$, $X_{\rm fl}\sim -\gamma\sim \dt^{(4-d)/2}$ decays to
zero.  
Conversely, if $V_\chi>V_C$ the main axis of the ellipse is along
the $\hat\chi$ direction and the fluctuation slope $X_{\rm fl}\sim -1/\gamma \sim
\dt^{(d-4)/2}$ becomes vertical for $\dt\to\infty$. In the limiting case
where $V_C$ and $V_\chi$ are equal asymptotically,
$(V_C/V_\chi)^{1/2}-(V_\chi/V_C)^{1/2}$ in\eq{X_fl} would decay as
$\dt^{(4-d)/2}$, being controlled by the leading corrections to $V_C$; because
this is proportional to $\gamma$, a finite asymptotic value of $X_{\rm
fl}$ would result. 
From the 
high-$T$ expansion of Sec.~\ref{sec:high_T} 
one can see that at order $1/T^2$, differences between the asymptotic values
of $V_C$ and $V_\chi$ appear. These suggest that the scenario that should 
apply is the one where $V_\chi>V_C$ asymptotically and therefore 
the main axis of the fluctuation ellipse is along
the $\hat\chi$ direction.

\subsection{Non-equilibrium, $d>4$} 
\label{sec:d_gt_4_nonequilibrium}

Now we study the behaviour of correlation and susceptibility
fluctuations for the genuine
out-of-equilibrium dynamics after a quench to criticality, focussing
on $d>4$ as in the previous subsection.
Here we get correction factors with respect to the 
equilibrium case 
which become important in the aging regime, $t-\tw\sim\tw$. 
More precisely, aging effects appear in the long time ($t,\tw\gg 1$)
behaviour of two-time functions via scaling functions of the time
ratio $t/\tw$ that modulate
the equilibrium part, \eg
\be
K(t,\tw)=K_{\rm eq}(t-\tw)\sc{K}\!\left(\textstyle\frac{t}{\tw}\right)
\label{K_noneq}
\ee
\be
\Ltwo(t,\tw)=\Ltwo\eql(t-\tw)\sc{L}\!\left(\textstyle\frac{t}{\tw}\right)
\label{L_noneq}
\ee
with $\sc{K}(1)=\sc{L}(1)=1$~\cite{AnnSol06}. Here and throughout the
remainder of the paper we distinguish the TTI equilibrium
contributions with the subscript ``eq''. 
The two-time correlation $C(t,\tw)$ has a similar scaling behaviour as
can be seen by expressing it in terms of the kernel $K$:
\bea
\fl C(t,\tw)&=&\dq R_{\qv}(t,\tw)C_{\qv}(\tw,\tw)
=\sqrt{\frac{g(\tw)}{g(t)}} \dq e^{-\om(t-\tw)}C_{\qv}(\tw,\tw)\\
\fl &=&\frac{g(\bar{t})}{\sqrt{g(t)g(\tw)}}K(\bar{t},\tw)
=
K\eql\!\left(\textstyle\frac{\dt}{2}\right)
\frac{g(\bar{t})}{\sqrt{{g(t)}g(\tw)}}\sc{K}\!\left(\textstyle\frac{\bar{t}}{\tw}\right)
\eea
The long-time behaviour of the function $D$ from\eq{D} that defines
the correlation variance is thus given by (using \eq{D_pieces},\eq{D1}
and\eq{D2})
%
\bea
\fl D(t,\tw,t')&=&\half
K\eql\!\left(\textstyle\frac{\dt}{2}\right)\frac{g(\bar{t})}{\sqrt{{g(t)}g(\tw)}}
\sc{K}\!\left(\textstyle\frac{x+1}{2}\right)[\delta(t-t')+\delta(\tw-t')]\nn
\fl &&{}+D_1(t,\tw,t')\theta(t'-\tw)+D_2(t,\tw,t')\theta(\tw-t')
\label{D_pieces_noneq}
\eea
where $x=t/\tw$ and
\be
\fl D_1(t,\tw,t')=\half K\eql\!\left(\textstyle\frac{\dt}{2}\right)
\frac{g(\bar{t})}{\sqrt{{g(t)}g(\tw)}}
\sc{K}\!\left(\textstyle\frac{x+1}{2}\right)\left[2T-\int_{t'}^t d\tau\,
L^{(2)}\eql(\tau-t')\sc{L}\!\left(\textstyle\frac{\tau}{t'}\right)\right]
\label{D1_noneq}
\ee
and
\bea
\fl D_2(t,\tw,t')&=&-\half K\eql\!\left(\textstyle\frac{\dt}{2}\right)
\frac{g(\bar{t})}{\sqrt{{g(t)}g(\tw)}}
\sc{K}\!\left(\textstyle\frac{x+1}{2}\right)\int_{\tw}^t d\tau\,
L\eql^{(2)}(\tau-t')\sc{L}\!\left(\textstyle\frac{\tau}{t'}\right)\nn
\fl &&{}+\frac{g(\bar{t})}{\sqrt{g(t)g(\tw)}}
\int_{\tw}^{\bar{t}} d\tau\, K\eql\!\left(\textstyle\bar{t}-\tau\right)\sc{K}\!\left(\textstyle\frac{\bar{t}}{\tau}\right)
L\eql^{(2)}(\tau-t')\sc{L}\!\left(\textstyle\frac{\tau}{t'}\right)
\label{D2_noneq}
\eea
For $d>4$ and long times $g(t)\sim {\rm const}$ and all the factors
involving $g(t)$ in the above expressions can be dropped.

In order to get the long-time expression for 
$D^{\chi}$ from\eq{Dchi_pieces} we need
the non-equilibrium form of $\chi(t,\tw)$. In equilibrium,
$1-T\chi\eql(\dt) = K\eql(\dt/2)$ and so we consider the same
combination here:
\be
1-T\chi(t,\tw)=1-T\dq\int_{\tw}^t dt'\,e^{-\om(t-t')}\sqrt{\frac{g(t')}{g(t)}}
\ee
This can be rewritten, by adding and subtracting the quantity
$T\dq\int_{\tw}^t dt'\,e^{-\om(t-t')}=1-K\eql(\dt/2)$, as
\be
1-T\chi(t,\tw)=K\eql\left(\textstyle\frac{\dt}{2}\right)
-T\dq\int_{\tw}^t dt'\,e^{\om(t-t')}\left(\sqrt{\frac{g(t')}{g(t)}}-1\right)
\ee
and extracting a factor of $K\eql(\dt/2)$ from the second term yields
\be
1-T\chi(t,\tw)=K\eql\!\left(\textstyle\frac{\dt}{2}\right)\sc{\chi}(x)
\label{chi_scaling}
\ee
Bearing in mind that $\dq e^{-\om(t-t')}=-K\eql'(t-t')/(2T)$ the
aging function $\sc{\chi}$ can be written as 
\bea
\sc{\chi}(x)&=&1+\half\frac{\int_{\tw}^t dt'\,
K\eql'\left(\textstyle\frac{t-t'}{2}\right)
\left(\sqrt{\frac{g(t')}{g(t)}}-1\right)}{K\eql(\dt/2)}
\label{Fchi_generic}
\eea
Because $g(t)$ approaches a constant for large times in $d>4$, the
integral term in fact vanishes in the limit and there is no aging correction:
$\sc{\chi}(x)=1$. Inserting\eq{chi_scaling}
into\eq{Dchi_pieces},\eq{D1chi} and\eq{D2chi}  
one finds the long-time non-equilibrium form of $D^{\chi}$
\bea
\fl D^{\chi}(t,\tw,t')&=&
\left[1\!-\!K\eql\!\left(\textstyle{\frac{\dt}{2}}\right)\sc{\chi}(x)\right]\delta(t-t')\nn
\fl & &\!+\! 
D_1^{\chi}(t,\tw,t')\theta(t'-\tw)\!+\!D_2^{\chi}(t,\tw,t')\theta(\tw-t')
\label{Dchi_pieces_noneq}
\eea
with
\bea
\fl D_1^{\chi}(t,\tw,t')&=&
\frac{1}{2}K'\eql\!
\left(\textstyle\frac{t-t'}{2}\right)
\sc{\chi}\!\left(\textstyle\frac{t}{t'}\right)
+\frac{t}{t'^2}K\eql\!\left(\textstyle\frac{t-t'}{2}\right)
\sc{\chi}'\!\left(\textstyle\frac{t}{t'}\right)
\nn
\fl &&{}+
2T\left[K\eql\!\left(\textstyle\frac{t-t'}{2}\right)\sc{\chi}(\textstyle\frac{t}{t'})-K\eql\!\left({\textstyle\frac{\dt}{2}}\right)
\sc{\chi}(x)\right]
\nn
\fl&&{}-\int_{t'}^t dt''\,L\eql^{(2)}(t''-t')\sc{L}(\textstyle\frac{t''}{t'})
\left[K\eql\!\left(\textstyle\frac{t-t''}{2}\right)\sc{\chi}(\textstyle\frac{t}{t''})-K\eql\!\left(\textstyle\frac{\dt}{2}\right)\sc{\chi}(x)\right]
\label{D1chi_noneq}
\eea
and
\be
\fl D_2^{\chi}(t,\tw,t')=
-\int_{\tw}^t dt''\,L\eql^{(2)}(t''-t')\sc{L}(\textstyle\frac{t''}{t'})
\left[K\eql\!\left(\textstyle\frac{t-t''}{2}\right)\sc{\chi}(\textstyle\frac{t}{t''})-K\eql\!\left(\textstyle\frac{\dt}{2}\right)\sc{\chi}(x)\right]
\label{D2chi_noneq}
\ee
It will be useful to separate
short and long time parts in $D^\chi$ again:
\be
D^\chi(t,\tw,t') = D^\chi_{\rm short}(t,\tw,t') + D^\chi_{\rm long}(t,\tw,t')
\label{Dchi_short_long_decomposition_again}
\ee
We arrange the terms so that the first term is identical to its
equilibrium counterpart\eq{Dshort} except for a slowly varying aging
correction, giving
\bea
\fl D_{\rm short}^{\chi}(t,\tw,t')&=&
\sc{\chi}\!\left(\textstyle\frac{t}{t'}\right)
\biggl[\delta(t-t') + \frac{1}{2}K'\eql\!
\left(\textstyle\frac{t-t'}{2}\right) +
2T K\eql\!\left(\textstyle\frac{t-t'}{2}\right)
\nn
\fl& &{}- \int_{t'}^t dt''\,L\eql^{(2)}(t''-t')
K\eql\!\left(\textstyle\frac{t-t''}{2}\right)\biggr]\theta(t'-\tw)
\label{D1chi_short_noneq}
\\
\fl D^\chi_{\rm long}(t,\tw,t') &=& {}-
K\!\left(\textstyle\frac{\dt}{2}\right)\sc{\chi}(x)\delta(t-t') +
D^{\chi}_{1,\rm long}(t,\tw,t')\theta(t'-\tw)\!
\nn
\fl &&{}+\!D_2^\chi(t,\tw,t')\theta(\tw-t')
\label{Dchi_long_noneq}
\\
\fl D_{1,\rm long}^{\chi}(t,\tw,t')&=&
\frac{t}{t'^2}K\eql\!\left(\textstyle\frac{t-t'}{2}\right)
\sc{\chi}'\!\left(\textstyle\frac{t}{t'}\right)
-K\eql\!\left(\textstyle\frac{\dt}{2}\right)\!\sc{\chi}(x)\!
\left[2T-\int_{t'}^t
dt''\,L\eql^{(2)}(t''\!-\!t')
\sc{L}\!\left(\textstyle\frac{t''}{t'}\right)\right]
\nn
\fl&&{}-\int_{t'}^t dt''\,L\eql^{(2)}(t''-t')
K\eql\!\left(\textstyle\frac{t-t''}{2}\right)
\left[\sc{L}\!\left(\textstyle\frac{t''}{t'}\right)
\sc{\chi}\!\left(\textstyle\frac{t}{t''}\right)
-\sc{\chi}\!\left(\textstyle\frac{t}{t'}\right)\right]
\label{D1chi_long_noneq}
\eea
We have kept the factors of $\sc{\chi}$ so that the expressions
are valid also in $d<4$, for later use. For our current case ($d>4$),
one has $\sc{\chi}=1$ and $\sc{\chi}'=0$.

To deduce the behaviour of $V_C$, $V_\chi$ and $V_{C\chi}$ we finally
need the aging corrections to the function $CC$ defined
in\eq{CC2times},\eq{CC3times}. One uses\eq{C_twotime} and the
long-time scaling of the equal-time correlator
$C_{\qv}(\tw,\tw)=(T/\om)\sc{C}(\om \tw)$ with $\sc{C}(w)\rightarrow
1$ for $w\rightarrow \infty$ and $\sc{C}(w)\approx 2w$ for
$w\rightarrow 0$; in $d>4$,
$\sc{C}(w)=1-\exp(-2w)$~\cite{AnnSol06}. The ratio of the three-time
function to its 
equilibrium counterpart is then, for $t'>\tw$,
\be
\frac{CC(t,\tw,t')}{CC\eql(t,\tw,t')} =
\frac{\dq (T^2/\om^2) \sc{C}(\om t')\sc{C}(\om\tw) e^{-\om(t-\tw)}}
{\dq (T^2/\om^2) e^{-\om(t-\tw)}}
\ee
In the aging regime where $t-\tw\gg 1$ and so only $\om\ll 1$
contributes one can replace $(dq) \sim d\omega\, \omega^{d/2-1}$;
rescaling $\om$ to $w=\om\tw$ then gives a function of $x=t/\tw$ and
$y=t'/\tw$:
\be
\fl \frac{CC(t,\tw,t')}{CC\eql(t,\tw,t')} = \sc{CC}(x,y), \quad
\sc{CC}(x,y)=\frac{\int dw\, w^{(d-6)/2} \sc{C}(wy)\sc{C}(w) e^{-w(x-1)}}
{\int dw\, w^{(d-6)/2} e^{-w(x-1)}}
\label{CC_scaling_tprime_gt_tw}
\ee
With the same approach one finds for $t'<\tw$, \ie\ $y<1$,
\be
\sc{CC}(x,y)=\frac{\int dw\, w^{(d-6)/2} \sc{C}^2(wy) e^{-w(x+1-2y)}}
{\int dw\, w^{(d-6)/2} e^{-w(x+1-2y)}}
\ee
The function $\sc{CC}$ also governs the aging corrections 
for the two-time $CC$, according to
\be
\frac{CC(t,\tw)}{CC\eql(t-\tw)} = 
\frac{CC(t,t,\tw)}{CC\eql(t,t,\tw)} = \sc{CC}(1,1/x)
\ee

We can now proceed to study what, if any, aging corrections there are
for the correlation and susceptibility (co-)variances in $d>4$. The scaling
functions $\sc{K}$, $\sc{L}$, $\sc{\chi}$, $\sc{CC}$ are all bounded
by $1$. 
One can check that the presence of the aging corrections in $D$ and 
$D^{\chi}$ does not increase the order of the various terms and that the 
asymptotic values of $V_C$ and $V_{\chi}$ are identical to the ones in 
equilibrium, with aging corrections visible only very weakly in the decay 
to these asymptotes. A detailed discussion of the effects of aging corrections 
on $V_C$ and $V_{\chi}$ is provided in~\cite{AnnSol08}.
Here we mention only that the leading correction to the asymptotic value 
of $V_C$ displays a crossover from the value $CC\eql(t-\tw)$, obtained
in the near 
equilibrium regime $x\approx 1$ (where $\sc{CC}(1,1/x)\approx 1$), 
to an $x$-independent negative constant of order $\tw^{(4-d)/2}$ for $x\gg 1$.
This negative contribution provides a small downward shift in the value of
$V_C$ that is reached asymptotically, as $t$ becomes large ($t\gg\tw$)
at fixed $\tw$.
The only significant aging effect in critical coarsening for $d>4$
is visible in the covariance, whose 
leading term can be shown to be~\cite{AnnSol08}
$V_{C\chi}=-2CC\eql(\dt/2)\sc{CC}(x,x)\sim
-\tw^{(4-d)/2}(x-1)^{(4-d)/2}\sc{CC}(x,x)$, that is the
equilibrium result supplemented with an aging correction $\sc{CC}(x,x)$. For
$x\approx 1$ the latter equals one as it should to reproduce the
equilibrium result; for large $x$, one finds
from\eq{CC_scaling_tprime_gt_tw} that $\sc{CC}(x,y)\sim y/x^2$ and so
$\sc{CC}(x,x)\sim 1/x$.

\section{Quenches to $\Tc$, $d<4$}
\label{sec:d_lt_4}

In this section we will study the fluctuations in the
out-of-equilibrium dynamics 
of the spherical ferromagnet quenched to criticality in dimension $d<4$. 
Here we cannot start from an equilibrium calculation and later account for
aging corrections because a naive equilibrium limit leads to
the appearance of infinite terms: in the correlation
variance\eq{VC_D}, for example, the first Gaussian term becomes $\dq
T^2/\om^2$ for $t,\tw\to\infty$ which is infinite in $d<4$.
Thus, we need to look directly at the non-equilibrium
situation. Specifically, we will consider the aging limit
$t,\tw\to\infty$ but at fixed $x=t/\tw>1$.
We omit the calculations for this scenario as they are somewhat
technical; the interested reader can find full derivations in
Ref.~\cite{AnnSol08}.

Our analysis shows that in the aging regime
the variances and the covariance all scale as 
$\tw^{(4-d)/2}$ times a function of $x$. The dependence on $x$ implies
that the relevant 
timescale on which the (co-)variances vary is $\dt\sim \tw$, in
contrast to the case $d>4$ where this timescale is $\dt=\order(1)$
independently of the age $\tw$. In this sense the behaviour for $d<4$ is
similar to what is seen for coarsening (in general $d$) at
$T<\Tc$~\cite{MayBisBerCipGarSolTra04,MaySolBerGar05}. Interestingly, however, the
amplitude of the (co-)variances grows with the age $\tw$ only as
$\tw^{(4-d)/2}$ at criticality, whereas below $\Tc$ it scales (at
least for $V_C$)
in the naive way as the domain volume $\sim(\tw^{1/2})^d$.

\begin{figure}
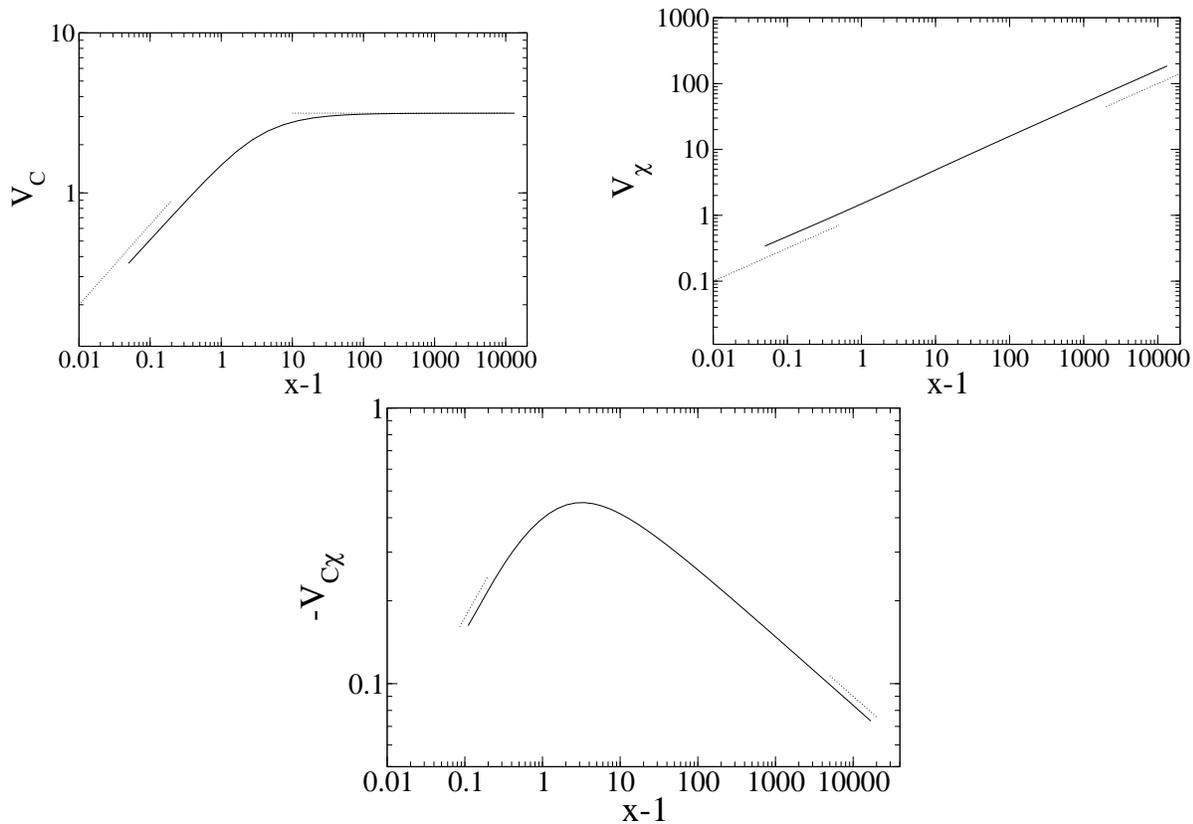

\includegraphics[width=7.0cm,clip=true]{0708corr3.eps}
\includegraphics[width=7.7cm,clip=true]{0708respd3.eps}
\centerline{\includegraphics[width=8.0cm,clip=true]{1008cross2d3.eps}}
\caption{Log-log plots of correlation and susceptibility variances
$V_C$ (top left), $V_\chi$ (top right) and covariance $V_{C\chi}$
(bottom) versus $x-1=(t-\tw)/\tw$, for critical coarsening   
in dimension $d=3$. As in all following figures, the (co-)variances have
been divided by $\tw^{(4-d)/2}$ to get functions of $x$ only.
The theoretically expected power laws
for large $x$ are given, respectively, 
by $V_C\sim x^0$, $V_\chi \sim x^{(4-d)/2}$ and $V_{C\chi}\sim
-x^{2-3d/4}$ and are represented by the dotted lines on the right of each
graph. 
The $\order{(1)}$ value approached by the correlation variance 
is represented by the Gaussian term and it can be calculated 
analytically~\cite{AnnSol08};
this value 
is represented by the horizontal dotted line in the first graph.
The initial increase $\sim(x-1)^{(4-d)/2}$ for all three quantities is
indicated similarly on the left. Note that for $V_\chi$ the initial
and asymptotic power laws are identical, so that the log-log plot
overall is close to linear. }
\label{fig:var_genericX3}
\end{figure}
\begin{figure}
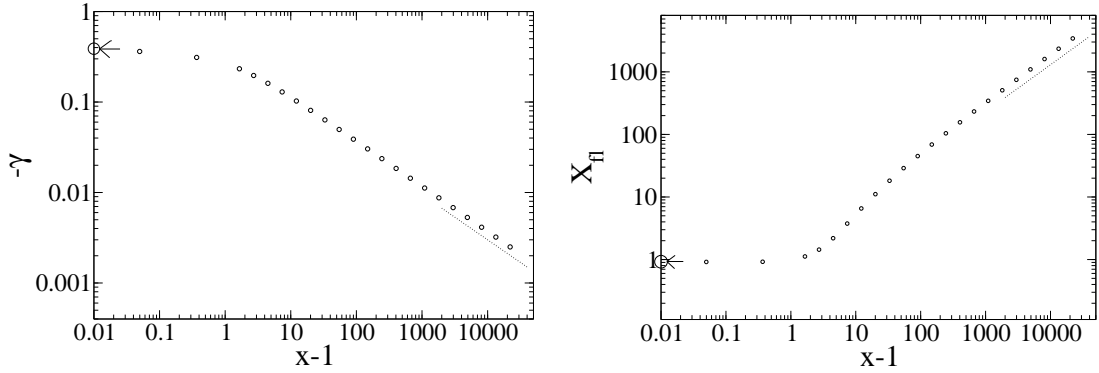

\centerline{\includegraphics[width=7.0cm,clip=true]{0708gammad3.eps}
\hspace{0.2cm}
\includegraphics[width=7.0cm,clip=true]{0708Xfld3.eps}}
\caption{Log-log plot of the absolute value of the 
correlation coefficient $\gamma$ (left) 
and the fluctuation slope $X_{\rm fl}$ (right) versus $x-1=(t-\tw)/\tw$ for
critical coarsening in $d=3$.
The modulus of the correlation coefficient $\gamma$ is always smaller
than $1$,
as it should be. For large $x$, $\gamma$ decays as $-x^{(2-d)/2}$, as
shown by the dotted line on the right of the plot. 
The separately calculated limit value
for $x\rightarrow 1$ is indicated by the
arrow on the $y$-axis and is certainly plausible as an asymptote for
$\ln(x-1)\to-\infty$ of our numerics for
$x$ close to 1.
The fluctuation slope $X_{\rm fl}$ 
behaves for large $x$ as $x^{d/4}$. This predicted power law is 
represented
by the dotted line on the right, while the value that should be approached 
for $x\rightarrow 1$ is indicated by the arrow on the $y$-axis.
}
\label{fig:gamma_Xfl_genericX}
\end{figure}

The dependence on $x$ of the (co-)variances can be evaluated
numerically from the scaling expressions for the aging
limit; the results are shown 
in Fig.~\ref{fig:var_genericX3} for $d=3$.
We also study numerically the $x$-dependence 
of the resulting correlation coefficient $\gamma$ and the fluctuation slope
$X_{\rm fl}$, as displayed in 
Fig.~\ref{fig:gamma_Xfl_genericX}.
In both of these quantities
the $\tw^{(4-d)/2}$ prefactor from the (co-)variances cancels, so they
depend solely on $x$.
\begin{figure}
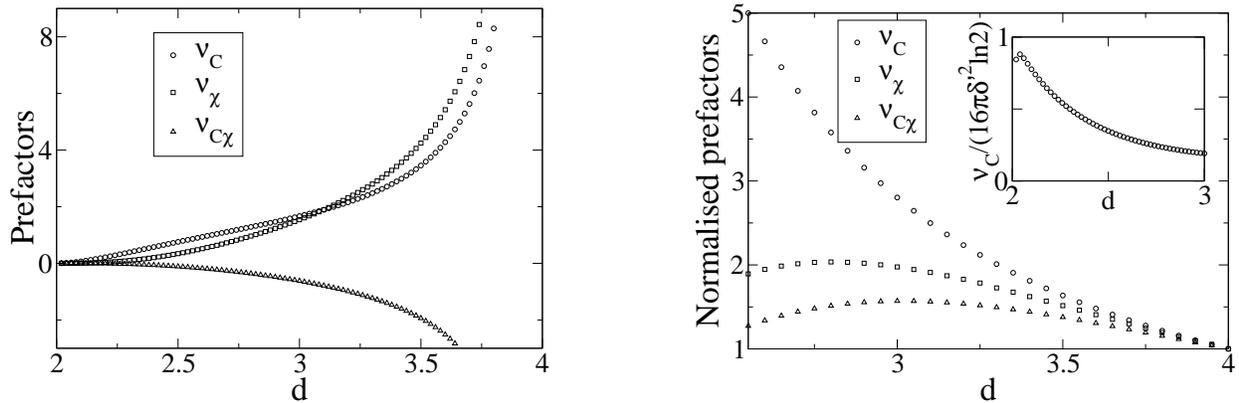

\setlength{\unitlength}{0.40mm} 
\begin{picture}(200,155)(-100,10)
\put(-103,10){\includegraphics[width=180\unitlength]{Var_x1PS.eps}}
\put(125,10){\includegraphics[width=180\unitlength]{NormalisedPrefactorsPS.eps}}\end{picture}
\caption{Left: The prefactors of the (co-)variances in the TTI regime
$x=t/\tw\approx 1$, where they are proportional to
$(t-\tw)^{(4-d)/2}$, are plotted versus $d$.
All three prefactors diverge as $1/(4-d)$ for 
$d$ close to $4$ and vanish as power
 laws
for $d$ near $2$, specifically 
$\nu_C\sim (d-2)^2$ and $\nu_{\chi}\sim \nu_{C\chi}\sim(d-2)^3$.
Right: 
Plot of the prefactors normalised by their predicted limiting behaviour 
as $\delta=(4-d)/2\to 0$~\cite{AnnSol08},
$16 \pi^2 \delta \nu_C/(3 \Tc^2)$, $4\pi^2\delta\nu_{\chi}/\Tc^2$ 
and $-8\pi^2\delta\nu_{C\chi}/\Tc^2$, versus $d$. The inset shows the 
correlation variance prefactor normalized by its predicted limiting behaviour 
as $\delta'=(d-2)/2\to 0$. 
}
\label{fig:var_x1}
\end{figure}
\begin{figure}
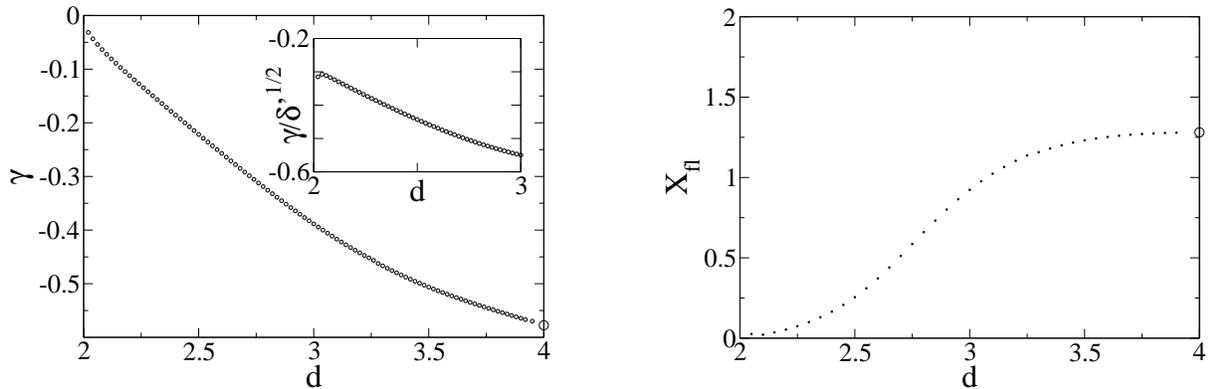

\setlength{\unitlength}{0.40mm} 
\begin{picture}(200,155)(-100,10)
\put(-103,10){\includegraphics[width=180\unitlength]{pgammaPS.eps}}
\put(115,10){\includegraphics[width=180\unitlength]{Xfl1.eps}}
\end{picture}
\caption{Correlation coefficient $\gamma$ and fluctuation slope $X_{\rm fl}$ 
for $x=t/\tw\rightarrow 1$, plotted versus $d$. The circles show the 
limits of $\gamma$ and $X_{\rm fl}$ for $d\rightarrow 4$, 
calculated analytically and given respectively 
by $-1/\sqrt{3}$ and $(1+\sqrt{17})/4$ (see~\cite{AnnSol08}). 
The inset on the left shows 
the correlation coefficient 
normalized by its predicted power law as it approaches $d=2$. The loss of 
accuracy around $d=2$ is due to numerical inaccuracies in the evaluation 
of the correlation variance as $d$ gets very close to $2$~\cite{AnnSol08}. 
}
\label{fig:gamma_Xfl_x1}
\end{figure}

One can analyse the behaviour of the quantities above in more detail
for the two 
opposite extremes $\epsilon=x-1\ll 1$ and $x\gg 1$. In the former case
we expect to recover quasi-equilibrium behaviour with all dependences
being only on time differences. One can show by an expansion in $\epsilon$~\cite{AnnSol08}
that indeed the (co-)variances all grow 
as $\epsilon^{(4-d)/2}$ to leading order; combining this with the
overall $\tw^{(4-d)/2}$ scaling, one gets a TTI time
dependence as expected, proportional to $(t-\tw)^{(4-d)/2}$.
The prefactors $\nu_C$, $\nu_{\chi}$ and $\nu_{C\chi}$
of this power law increase in $V_C$, $V_\chi$ and $V_{C\chi}$
are plotted as functions of dimensionality in Fig.~\ref{fig:var_x1}.
Given that $V_C$, $V_\chi$, $V_{C\chi}$ all have the same scaling with
$t-\tw$ in this regime, the
correlation coefficient and fluctuation slope also have nontrivial
values; these are plotted against $d$ in
Fig.~\ref{fig:gamma_Xfl_x1}. Note that because we are approaching the
TTI regime via the limit $x\to 1$ of an aging calculation, where
$t-\tw\sim \tw\gg 1$ always, these results are valid for $t-\tw\gg 1$
only. For $t-\tw\ll 1$, on the other hand, the results of
Sec.~\ref{sec:short_time_differences} will apply and both $\gamma$ and
$X_{\rm fl}$ will tend to zero in the limit $t\to\tw$. If we were to
plot the contour lines of the distribution of $(\hat C,T\hat\chi)$ on
an FD plot the initial section (bottom right) would therefore
look similar to Fig.~\ref{fig:FDHighT}, but then the ellipses would
grow as $(t-\tw)^{(4-d)/2}$ as the top left hand corner of the plot is
approached and their principal axis would approach a limiting slope
given by $X_{\rm fl}$ in Fig.~\ref{fig:gamma_Xfl_x1} as a function of
$d$. The genuine aging effects occurring for $x>1$ would not be
visible because they are all compressed into the top left corner of
the plot in the limit $\tw\to\infty$.

Analytically one can obtain relatively simple expressions for the
prefactors $\nu_C$, $\nu_\chi$, $\nu_{C\chi}$
in the limits $d\rightarrow 2$ and $d\rightarrow 4$~\cite{AnnSol08}.
We mention here only that they all vanish as power
laws of $d-2$ in the limit $d\to 2$. 
This makes sense intuitively in
that for $d<2$ no phase ordering takes place and so the fluctuations
arising from the coarsening dynamics should vanish as $d\to 2$ from
above. 

Turning to the opposite limit of large $x$, 
we find~\cite{AnnSol08} that the correlation variance $V_C$
is dominated by the Gaussian term
asymptotically, matching qualititatively the behaviour in the regime
$d>4$.
Here, however, the first subleading correction
to the constant asymptote is negative, so that the approach is from
below, in contrast to $d>4$ (see Fig.~\ref{fig:var_genericX3}). 
Quantitatively
the correction term is small already in $d=3$ and numerical evaluation
shows that is gets progressively smaller as $d$ increases to $4$.
For the response variance and the covariance
we get
$V_{\chi}\sim x^{(4-d)/2}$ and $V_{C\chi}\sim -x^{2-3d/4}$. All of these
scalings, as well as those for $x\to 1$, are in agreement with our
numerical evaluations as shown in 
Fig.~\ref{fig:var_genericX3} above.

We discuss briefly the consequences of the above results for the
large-$x$ behaviour of the contour ellipses of the joint distribution
$P(\hat C,T\hat\chi)$ of the fluctuating correlation and susceptibility.
Firstly, due to the different scaling with $x$ of the correlation and
susceptibility variances, as 
$x^0$
and $x^{(4-d)/2}$ respectively, these ellipes become increasingly
elongated in the susceptibility direction as $x$ grows large.
Using also the scaling of the covariance $V_{C\chi}\sim -x^{2-3d/4}$,
the correlation coefficient $\gamma$ from\eq{corr_coef} 
decays to zero aymptotically as $-x^{(2-d)/2}$. 
Finally we consider the fluctuation slope $X_{\rm fl}$. 
Looking at\eq{X_fl}, one can check~\cite{AnnSol08} that
this is given by
$X_{\rm fl} = -2(V_{\chi}/V_C)^{1/2}/(2\gamma)=-V_\chi/V_{C\chi}\sim
x^{d/4}$. 
The large-$x$ divergence of $X_{\rm fl}$ as $x^{d/4}$ for any $2<d<4$ is 
consistent with the fact that 
the joint distribution of correlation and response fluctuations grows
more quickly in the susceptibility direction than along the
correlation axis. 
It also matches our finite $x$ numerics, 
see Fig.~\ref{fig:gamma_Xfl_genericX}.

\section{Discussion}
\label{sec:discussion}

We have analysed fluctuations in the coarsening dynamics of the
spherical ferromagnet after a quench, specifically the leading $1/\sqrt{N}$
fluctuations of local correlations and susceptibilities spatially
coarse-grained across the entire system. Our work was inspired by
general theories regarding the nature of correlation and
susceptibility fluctuations in aging
systems~\cite{CasChaCugKen02,ChaKenCasCug02,CasChaCugIguKen03,ChaCug07}.
Our study significantly extends the scope of previous
(zero-temperature) calculations of correlation
fluctuations in the spherical model~\cite{ChaCugYos05} by keeping track of non-Gaussian
fluctuations. This enables us to calculate explicitly the susceptibility
fluctuations, which in a Gaussian approximation would vanish
identically. The nature of our approach, which treats the non-Gaussian
effects perturbatively, means that we cannot analyse quenches to below
the critical temperature; however, we {\em can} access the interesting
regime of coarsening {\em at} criticality, where our results are the
first of their kind.

We discussed carefully in Sec.~\ref{sec:definitions} possible definitions of
coarse-grained fluctuating correlations $\hat C$ and responses
$\hat\chi$. It turns out that for the fluctuation statistics (in
non-disordered systems such as the one studied here) it does matter
whether the underlying local functions are measured directly, or
indirectly via quenched amplitudes that define a randomly staggered
magnetization observable: only the former choice gives correlation and
susceptibility variances that scale in the same way with system size
$N$. These considerations should be of general relevance to other
systems where a coarse-graining of the local correlation and response
across an entire finite-sized system is desired. Coarse-graining over
smaller volumes does not produce interesting results in the spherical
model because the susceptibility fluctuations are small ($\sim
N^{-1/2}$) but correlated across the entire system.

In Sec.~\ref{sec:setup_hetero} we used the $1/\sqrt{N}$ expansion of
the non-Gaussian fluctuations~\cite{AnnSol06} to derive general
expressions for the correlation and susceptibility variances and
covariance. These are exact to leading order in $1/N$, where the
joint distribution of $\hat C$ and $\hat \chi$ is Gaussian. 
In addition to $V_C$, $V_{\chi}$ and the covariance $V_{C\chi}$,
this distribution can be characterized by the correlation coefficient
$\gamma$ (see equation\eq{corr_coef}) and the negative slope of the
principal axis of the elliptical equi-probability contours, $X_{\rm
fl}$ (see equation\eq{X_fl}). The definition of $X_{\rm fl}$ was chosen
such that, if the predictions for glassy systems (such as spin
glasses) with a global time reparameterization
invariance~\cite{CasChaCugKen02,ChaKenCasCug02,CasChaCugIguKen03,ChaCug07}
applied also to coarsening systems, $X_{\rm fl}$ should be close to
the fluctuation-dissipation ratio (FDR) $X$ that relates the variations with time
of the {\em average} susceptibility and correlation.

In Sec.~\ref{sec:hetero_above_Tc} we considered first quenches to
$T>\Tc$, where after fast initial transients the dynamics is in
equilibrium. Analytical results were obtained in the limit of high
temperatures: here the probability contours rotate with increasing
time difference $\dt=t-\tw$ from a horizontal orientation ($X_{\rm fl}=0$) to 
a vertical one ($X_{\rm fl}=\infty$).
At the
same time the correlations between correlation and susceptibility
fluctuations become weaker and weaker: the contours become approximately
circular,
showing an effect opposite to the progressive narrowing of the
contours around the slope of the fluctuation dissipation plot ($X_{\rm
  fl}=1$ for large $T$)
that would be expected for spin glasses and similar
systems~\cite{CasChaCugKen02,ChaKenCasCug02,CasChaCugIguKen03,ChaCug07}.
Qualitatively this behaviour remains the same also for quenches to
lower temperatures above $\Tc$; the results for small time differences
$\dt$, in particular, depend on $T$ only through
prefactors. The correlation coefficient $\gamma$ between the
fluctuations of correlations and susceptibilities is always negative,
corresponding to a positive fluctuation slope $X_{\rm fl}$.

The more interesting quenches to criticality were studied in
Sec.~\ref{sec:hetero_dgt4} (for dimension $d>4$) and Sec.~\ref{sec:d_lt_4}
(for $d<4$). In the former case, we found that out-of-equilibrium effects
are weak and one can directly analyse the equilibrium dynamics.
Interestingly, the correlation variance -- which is identical to the
four-point correlation function often used to characterize dynamic
heterogeneities -- displays a maximum as a function of $\dt$,
suggesting as in other glassy systems that there is a well-defined timescale on
which fluctuations between different dynamical trajectories of a
system are largest. However, even though the coarsening dynamics has a
growing lengthscale that increases with the system age $\tw$ in the
standard diffusive manner, $\xi(\tw)\sim\tw^{1/2}$, neither the timescale of the maximum in $V_C$ nor its amplitude change with age. This is
in contrast to the case of coarsening below $\Tc$, where timescales
grow with the age and the variance has the natural scaling with $\xi^d(\tw)
\sim \tw^{d/2}$~\cite{MayBisBerCipGarSolTra04}.

Below $d<4$ one has to look directly at the non-equilibrium situation:
a naive equilibrium limit yields infinities that need to be
regularized by initially keeping the age $\tw$ finite. We found that
$V_C$, $V_\chi$ and $V_{C\chi}$ all scale as $\tw^{(4-d)/2}$ times
functions of the time ratio $x=t/\tw$.
Looking at the details of
the $x$-dependence, we saw that in the quasi-equilibrium regime
$x\approx 1$ time-translation invariance is restored as expected, with
all (co-)variances scaling as
$\tw^{(4-d)/2}(x-1)^{(4-d)/2}=(t-\tw)^{(4-d)/2}$ for $t-\tw\gg
1$. Unlike the case $d>4$, the correlation coefficient is finite in
this regime, but the corresponding fluctuation slope $X_{\rm fl}$ does
not seem to be related to the FDR -- which is $X=1$ at
quasi-equilibrium -- in any simple way. In particular, $X_{\rm fl}$
grows monotonically from a vanishing value at $d=2$ to the non-trivial limit
$(1+\sqrt{17})/4$ in $d=4$. In 
the genuine aging behaviour that follows
for larger $x$ 
the correlation coefficient 
decays to zero 
and the
fluctuation slope $X_{\rm fl}$ 
diverges towards
large
positive values as $x$ grows, both reflecting the progressive
stretching of the probability contours along the susceptibility axis.

From a more general point of view, our results show clearly that
heterogeneities are present in coarsening at criticality 
{\it above} the upper critical dimension,
as detected
\eg\ via maxima in $V_C$ in $d>4$. However, the lack of a dependence
on age $\tw$ in the relevant
timescales ($\dt=\order(1)$) 
and amplitudes
($V_C=\order(1)$) 
is somewhat surprising. Interestingly, the maximum in $V_C$ is seen to
disappear {\em below} the critical dimension. The timescale on which $V_C$
varies then has a conventional aging form ($\dt=\order(\tw)$) while
its amplitude $V_C=\order(\tw^{(4-d)/2})$ is not related in any
obvious manner to the growing correlation volume of order $\tw^{d/2}$.

It will be an interesting
challenge to see whether the general features of our results, and in
particular the different $\tw$-scalings of the correlation variance
above and below $d=4$, can be understood from general scaling
or field theoretical approaches to critical
coarsening~\cite{CalGam05}. One would also like to extend our
considerations to genuinely short-ranged systems: the spherical model
is somewhat unusual in that the spherical constraint generates a weak
but long-range interaction. The $\order(n)$ model in the limit of
large $n$ may be a suitable candidate here; preliminary work suggests
that much of our perturbative approach for analysing non-Gaussian
fluctuation effects would transfer to this scenario. Finally, it is
clear from our results that fluctuations in critical coarsening
display very rich behaviour that cannot simply be deduced from the
properties of the average fluctuation-dissipation relations, and it
remains to be seen whether alternative ways can be found of
rationalizing the kind of effects thrown up by our exact calculations.

\section*{References}

\bibliography{hetero_ref}
\bibliographystyle{unsrt}

\end{document}